# Tuning Sizes, Morphologies, and Magnetic Properties of Monocore Versus Multicore Iron Oxide Nanoparticles through the Controlled Addition of Water in the Polyol Synthesis

*Gauvin Hemery*[a], *Anthony C. Keyes Jr.*[a], *Eneko Garaio*[b]*, *Irati Rodrigo*[b,c], *Jose Angel Garcia*[c,d], *Fernando Plazaola*[b], *Elisabeth Garanger*[a], *Olivier Sandre*[a]*

[a] LCPO, CNRS UMR 5629/ Univ. Bordeaux/ Bordeaux-INP, ENSCBP 16 avenue Pey Berland, 33607 Pessac, France

[b] Elektrizitatea eta Elektronika Saila, UPV/EHU, 48940 Leioa, Spain

[c] BCMaterials, Parque Tecnológico de Bizkaia, Ed. 50, 48160 Derio, Spain

[d] Fisika Aplikatua II Saila, UPV/EHU, 48940 Leioa, Spain

KEYWORDS Polyol synthesis, iron oxide, nanoparticles, magnetic hyperthermia, MRI contrast agent, nanoflowers, multi-core, mono-core

ABSTRACT The polyol route is a versatile and up-scalable method to produce large batches of iron oxide nanoparticles with well-defined structures and magnetic properties. Importance of parameters such as temperature and reaction time, heating profile, nature of polyol solvent or of organometallic precursors on nanostructure and properties has already been described in the literature. Yet, the crucial role of water in the forced hydrolysis pathway has never been reported despite its mandatory presence for nanoparticle production. This communication investigates the influence of the water amount and temperature at which it is injected in the reflux system for either a pure polyol solvent system or a mixture with poly(hydroxy)amine. Distinct morphologies of nanoparticles were thereby obtained, from ultra-ultra-small smooth spheres down to 4 nm in diameter to larger ones up to 37 nm. Well-defined multi-core assemblies with narrow grain size dispersity termed nanoflowers were also synthesized. A diverse and large library of samples was obtained by playing with the nature of solvents and amount of added water while keeping all other parameters constant. The different morphologies lead to magnetic nanoparticles suitable for important biomedical applications such as magnetic hyperthermia, MRI contrast agent, or both.

## INTRODUCTION

Magnetic iron oxide nanoparticles (IONPs) find applications in the biomedical field as diagnostic tools and innovative therapies as they provide contrasting properties in magnetic resonance imaging (MRI) and also serve as heat mediators in magnetic fluid hyperthermia (MFH) and for triggered drug delivery.[1] As IONPs are biocompatible, they are among the best candidates over alternative magnetic nanomaterials for health applications. Various synthetic pathways have been explored for their production,[2] the most common ones being the alkaline co-precipitation,[3] the polyol process,[4] the thermal degradation of organometallic precursors,[5] and the hydrothermal treatment.[6] The polyol route is an interesting compromise when taking into account the ease of synthesis, scalability, cost-efficiency, and control over the morphology. This process leads to nanoparticles with morphologies ranging from smooth spheres to more complex structures such as the so-called "nanoflowers". Other strategies were reported to directly produce flower-like structures in water such as the co-precipitation performed in presence of excess polysaccharide,[7] or in a microwave reactor.[8] The polyol synthesis of multi-core IONPs was introduced by Caruntu *et al.*.[4] In this reaction pathway, the solvent acts simultaneously as a complexing agent for iron chloride precursors and as high boiling point solvent, with reflux temperatures usually in the order of 220 °C. Nanoflowers are to date amongst the best IONPs in terms of efficiency for heating under an applied alternating magnetic field (AMF) and as negative (transverse $T_2$) MRI contrast agents, as previously reported.[9] At equivalent concentrations of iron, the outstanding heating properties of these assemblies of small grains into larger raspberry-like structures were related to a frustrated super-spin glass state.[10] Regarding mechanistic pathway of the synthesis, tentative descriptions of the intermediary states of reaction were proposed,[11] together with the role of the polyol solvent in orienting the morphology.[12] Different reaction conditions were studied in the literature, with variable parameters such as the choice of the polyol solvent,[12] temperature and pressure,[13] reaction time and heating ramp slope,[14] alkaline pH,[15] and presence of adsorbed capping agents.[16] The main applications envisioned for nanoflowers are as nanoheaters for MFH and efficient negative ($T_2$) contrast agents for MRI, as ascribed to their large magnetic moment and large intrinsic magnetization $M_d$ (magnetic mono-domain moment divided by the particle volume). The main challenge to overcome for utilizing these superior magnetic properties in biological media is to prevent particle aggregation that can be evidenced by a non-reversible magnetization curve in static magnetic field. With large magnetic moments, nanoflowers experiment strong magnetic dipolar interactions. Moreover, the coating of their surface by a residual organic layer can render them difficult to peptize as stable aqueous colloidal suspensions, especially in physiological media. However, we found that all nanoparticles synthesized in our study, even the largest nanoflowers, show superparamagnetic behavior at room temperature with no hysteresis of their magnetization curve under DC field at room temperature, which is a prerequisite for their use in biological applications (yet still necessitating further coating to reach stealthiness in blood circulation).

In the course of our studies, we evidenced the crucial role of water in the polyol synthesis of IONPs. We report thereafter an extensive study describing how the amount and way of adding water in the synthesis influence the final nanoparticle (NP) morphology. A library of water-dispersible IONPs was therefore successfully synthesized, with tunable diameters from ~4 to ~37 nm and superior magnetic properties for MFH and MRI.

## EXPERIMENTAL SECTION

### Materials

Nitric acid ($HNO_3$, 70%) was from Fisher, ethyl acetate (>99.5%) was from Sigma Aldrich, acetone (technical grade), ethanol (96%), and diethyl ether (100%) were from VWR. *N*-methyldiethanolamine (NMDEA, 99%) was from Acros Organics, diethylene glycol (DEG, 99%), sodium hydroxide microprills (NaOH, 98%), iron(III) nitrate nonahydrate ($Fe(NO_3)_3 \cdot 9H_2O$, >98%), and iron(II) chloride tetrahydrate ($FeCl_2 \cdot 4H_2O$, 98%) were from Alfa Aesar. Iron(III) chloride hexahydrate ($FeCl_3 \cdot 6H_2O$, >97%) was from Panreac.

### Synthesis of iron oxide nanoparticles

Nanoparticles were produced by adapting reaction conditions reported by Caruntu *et al.*:[4] 80 mL of either pure DEG or a mixture of DEG and NMDEA with volume ratios 1:1 v/v were introduced in a three-neck round bottom flask flushed with nitrogen and stirred with a magnetic stir bar for one hour under inert atmosphere. 1.08 g (4 mmol) of $FeCl_3·6H_2O$ and 0.40 g (2 mmol) of $FeCl_2·4H_2O$ were then dissolved overnight. In the meantime, 0.64 g (16 mmol) of NaOH was dissolved under magnetic stirring in 40 mL of either pure DEG or a 1:1 v/v mixture of DEG and NMDEA in a separate three-neck round bottom flask. The NaOH solution was flushed by bubbling nitrogen for one hour before mixing with the mixed iron(II,III) chloride solution. The color quickly turned from yellow to deep green. The mixture was then heated up to 220 °C (temperature ramp in around 30 min) with an electronically controlled Digi-Mantle™ heating mantle (OMCA0250, Electrothermal™) set at full power, before letting the reaction to occur for a determined period of time, either with or without agitation at 500 rpm by a mechanical stirring Teflon shaft inserted through the condenser aperture. Nanoparticles were then separated over a strong permanent ferrite magnet ($152×101×25.4\ mm^3$, Calamit Magneti™, Milano-Barcelona-Paris), washed three times with a mixture of ethanol and ethyl acetate (1:1 v/v), once with 10 % nitric acid, twice with acetone, and twice with diethyl ether. NPs were then readily redispersed in water by stirring in open air to remove volatile solvents. At this stage, a black monophasic dispersion of IONPs was obtained. 8.6 g of iron(III) nitrate was then added as a strong oxidant by heating at 80 °C for 45 min while mechanically stirring.[17] The colloidal solution then turned from clear black to clear brown-orange. The IONPs were flocculated by addition of 10 % nitric acid and finally washed twice with acetone and twice again with diethyl ether. At this stage a deep orange-black dispersion of IONPs was obtained. The fluid was attracted by permanent magnets while staying in a single liquid phase, confirming that a true "ferrofluid" was obtained.

## Sample nomenclature

Each final product is designated according to the main synthesis parameters, *i.e.* the solvent (*D* for pure DEG, *N* for DEG/NMDEA 1:1), the volume of water in µL added to 120 mL of solvent (the subscript *HI* or *HU* being added to specify if water was injected to the reaction mixture at solvent reflux – hot injection – or by heating-up from room temperature, respectively), and the reaction time. For instance, the sample name $N500_{HU}$-$5h$ indicates that 500 µL of $H_2O$ were added to 120 mL DEG/NMDEA (1:1) and the reaction mixture was heated up to reflux for 5 hours. The sample name $D5000_{HI}$-$20m$ indicates that 5,000 µL of $H_2O$ were added through a septum to 120 mL of iron(II,III) precursors in boiling DEG, then let for 20 min before cooling).

**Table 1:** Batch names and their corresponding conditions of synthesis. Asterisk designates a reaction performed under 'natural mixing' *i.e.* by diffusion and convection yet no stirring.

| Batch name | Nomenclature |
|---|---|
| 15ff | $N1000_{HU}$-$5h$ |
| 17ff | $D5000_{HI}$-$20m$ |
| 25ff | $N500_{HU}$-$4h$ |
| 30ff | $N1000_{HU}$-$5h$ |
| 31ff | $N500_{HU}$-$1h$ |
| 32ff | $N500_{HU}$-$5h$ |
| 34ff | $N100_{HU}$-$5h$* |
| 35ff | $N100_{HU}$-$5h$ |
| 36ff | $N100_{HU}$-$5h$ |

## Transmission Electron Microscopy (TEM)

TEM was performed on a Hitachi™ H7650 microscope with an acceleration voltage of 80 kV. TEM images were acquired with an ORIUS™ SC1000 large format (11 MPx) camera. Samples were prepared by nebulizing NP dispersions at 1 $g{\cdot}L^{-1}$ concentration on Formvar™ carbon-coated 200 mesh copper grids from Agar Scientific™ and leaving them to dry at room temperature. NP size distributions were obtained by measuring more than 100 NPs with the ImageJ freeware (https://imagej.nih.gov/ij/). Size-histograms were fitted to a log-normal distribution law of diameters $P(d)$ with optimized values of median diameter $\alpha$ and non-dimensioned width $\beta$:

$$P(d) = \frac{1}{d \cdot \beta\sqrt{2\pi}} \cdot \exp\left(\frac{-(\ln(d) - \ln(\alpha))^2}{2\beta^2}\right)$$

In addition, the mean sizes $d_0 = <d>$ (number-averages) and standard deviations $\sigma = <(d-<d>)^2>^{1/2}$ were calculated using the classical Gaussian statistics formulas.

**Dynamic Light Scattering (DLS) and Zetametry**

A Nanosizer™ Nano ZS90 from Malvern™, UK, was used to measure $\zeta$ potentials, $Z$-average hydrodynamic diameters ($D_h$), and polydispersity indexes (PDI). The 2nd order Cumulant fit was used for analysing DLS data into a size distribution (the PDI being defined as the ratio of the 2nd order coefficient to the square of the 1st order one in the series[18]). The Smoluchowski equation was used to convert the measured electrophoretic mobility into a $\zeta$ potential value in mV.

**Attenuated total reflection infrared spectroscopy**

ATR-IR spectroscopy was carried on a GladiATR™ device from Pike Technologies mounted on a Bruker Vertex™ 70 FT-IR spectrometer. Typically a drop of sample was dried on the diamond prism using a hair blower followed by a 64 scans measurement. Spectra were analyzed with the Opus™ software.

### Proton relaxometry

Samples were prepared at concentrations of 6 $mM_{Fe}$. NMR tubes (7.5 mm outer diameter) were filled with 1 mL of each sample, and inserted in a Bruker™ mq60 relaxometer equipped with a 60 MHz / 1.41 Tesla magnet. The samples were left to thermalize to 37 °C using a Julabo™ f25 ED circulation bath. Following recommended protocols in proton relaxometry,[19] longitudinal $T_1$ relaxation times were measured using an inversion-recovery sequence of first duration of ~$0.1\times T_1$ and final duration of ~$3\times T_1$ with a recycling delay (RD) of ~$5\times T_1$ between two of the 4 acquisitions, 10 data points per scan, and an automatic RF receiver gain. Transverse $T_2$ relaxation times were measured using Car-Purcell-Meiboom-Gill (CPMG) sequence, with delay time $\tau$ of 0.04 ms between the 90° rotation to transverse plane and the 180° focusing pulse, a duration time of $3\times T_2$, RD of $5\times T_1$, and automatic receiver gain. The number of acquisition points was set by dividing the duration time by the delay time $\tau$.

### Magnetic fluid hyperthermia

NPs were dispersed at concentrations of 3 $g\cdot L^{-1}$ in diluted $HNO_3$ (at pH~2) to preserve their colloidal stability. The samples were placed in 500 µL plastic cuvettes, whose caps were pierced with a needle to introduce a fiber optics temperature probe of 420 µm outer diameter (medical range OTG-M420 fiber, Opsens™, Québec, QC, Canada) and measure temperature profiles *versus* time. Samples were thermalized at 37 °C using a glass-water jacket connected to a temperature bath until reaching equilibrium. The heat generation by magnetic NPs was triggered using an induction coil (4-turn of 3.5 mm diameter hollow – 0.4 mm wall – copper tubing, 55 mm outer diameter, 48 mm inner diameter, 34.5 mm height) fed by a Minimax Junior™ 1TS 3.5 kW generator (Seit Elettronica™, Italy) applying an alternating magnetic field (AMF) at maximum amplitude $H_{app}$ of 10.2 kA.m$^{-1}$ and at a frequency $f$ of 755 kHz as determined by finite element modelling.[20] The amplitude and frequency of the magnetic field were corroborated by measuring the electromotive force in a scout coil (made of a single turn of 17.5 mm diameter) and an oscilloscope (Agilent™ 54641 A). The AMF was applied for 5 min while recording the elevation of temperature and measuring its slope at early times (within first 5 s).

**Static (DC) magnetization**

DC magnetization curves of the NP aqueous dispersions were obtained up to 1.79 Tesla on a homemade vibrating sample magnetometer (VSM) at the SGIker facility (UPV/EHU). The magnetic field was measured by a gaussmeter whereas the signal was conditioned by a Stanford™ SR810DSP lock-in amplifier controlled by a PC under a LabVIEW™ program. All VSM measurements were performed at room temperature, the applied field $H_{app}$ being cycled from $1.43\times10^{6}$ A·m$^{-1}$ to $-1.43\times10^{6}$ A·m$^{-1}$ and then back to $1.43\times10^{6}$ A·m$^{-1}$.

**Zero field cooling – field cooling (ZFC-FC) magnetometry**

ZFC-FC experiments were conducted on a magnetic property measurement system (MPMS™ 7T from Quantum Design™, San Diego, CA, USA). This ultrasensitive magnetometer was previously calibrated by $Y_3Fe_5O_{12}$ garnet 1 mm diameter sphere (standard reference materials 2853) and reset after each measurement. Estimates of the blocking temperature were made according to a previously published protocol.[21, 22]

**Equivalent iron titration in IONP suspensions**

The equivalent iron molarity [Fe] was measured by a disruptive photometric assay, using the characteristic absorption peak at 350 nm of the $[Fe(Cl)_6]^{3-}$ complex when an aliquot of the suspension was dissolved in concentrated hydrochloric acid (HCl 5 M), according to previous calibration law $OD_{350nm,\ 2mm}=0.5043\times[Fe]_{mM}+0.0172$. Then [Fe] was converted into iron oxide weight assuming pure $\gamma$-$Fe_2O_3$ composition (~80 g per mol of iron).

**Small angle neutron scattering (SANS)**

SANS curves were acquired on the PACE spectrometer of the LLB-CEA Saclay, France, equipped with an isotropic $BF_3$ detector made of 30 concentric rings of 1 cm width each. Three configurations were used to cover overlapping $q$ ranges of $3.1\times10^{-3}$ – $3.3\times10^{-2}$, $1.4\times10^{-2}$ – $1.5\times10^{-1}$, and $4.7\times10^{-2}$ – 0.48 $Å^{-1}$, with the following values of sample-to-detector distance $D$ and neutron wavelength $\lambda$: $D$=4.57 m and $\lambda$=13 Å, $D$=2.85 m and $\lambda$=4.6 Å, $D$=0.87 m and $\lambda$=4.6 Å. After dividing scattered intensity by the transmission factor, subtracting the incoherent background, and normalizing by the flat signal of a cuvette filled with light water to correct the detector efficiency, the absolute intensity in $cm^{-1}$ was obtained.[23] Curve fitting by a polydisperse sphere form factor gave the $R_{SANS}$ radius and its standard deviation, and Porod's law at high $q$ knowing volume fraction $\phi$ gave an estimate of the specific area $A_{spe}$ (**Figure S1**).

**RESULTS AND DISCUSSIONS**

In the course of our work, we postulated that water traces are incorporated in the highly hygroscopic polyol solvent and that this initial water content in the mixture before reaction is a key parameter to produce well-defined magnetic IONPs. This led us to carry out an extensive study on the role of controlled additions of water in the synthesis of iron oxide nanoparticles in a mixture of polyols (DEG, NMDEA). Different batches of IONPs were thereby produced by varying reaction parameters such as reaction time, solvent system, amount and timing of injections of water amounts, *i.e.* hot injection at solvent reflux *vs.* heating-up of the reaction mixture including water. Since the two major biomedical applications of IONPs are for MFH and as MRI contrast agents, we have dedicated our efforts to produce well-defined IONPs suitable for these two applications. Large (several tens of nm) IONPs were produced for MFH, while ultra-ultra-small superparamagnetic iron oxide NPs (UUSPIO) of just a few nm were synthesized for use as positive MRI contrast agents with $T_1$-weighted sequences.[24-27]

**Study and optimization of the forced hydrolysis pathway in polyol**

The main reaction parameter to control IONP morphology (and therefore magnetic properties) was the solvent composition. A mixture of DEG and NMDEA (1:1 volume ratio) was used to yield large IONPs, while UUSPIO NPs were produced in pure DEG. In both cases, the same quantity and stoichiometric ratio of iron(II,III) chlorides and hydroxides were used ($Fe^{3+}/Fe^{2+}/OH^-$ 2:1:8, *i.e.* one hydroxide anion per chloride).

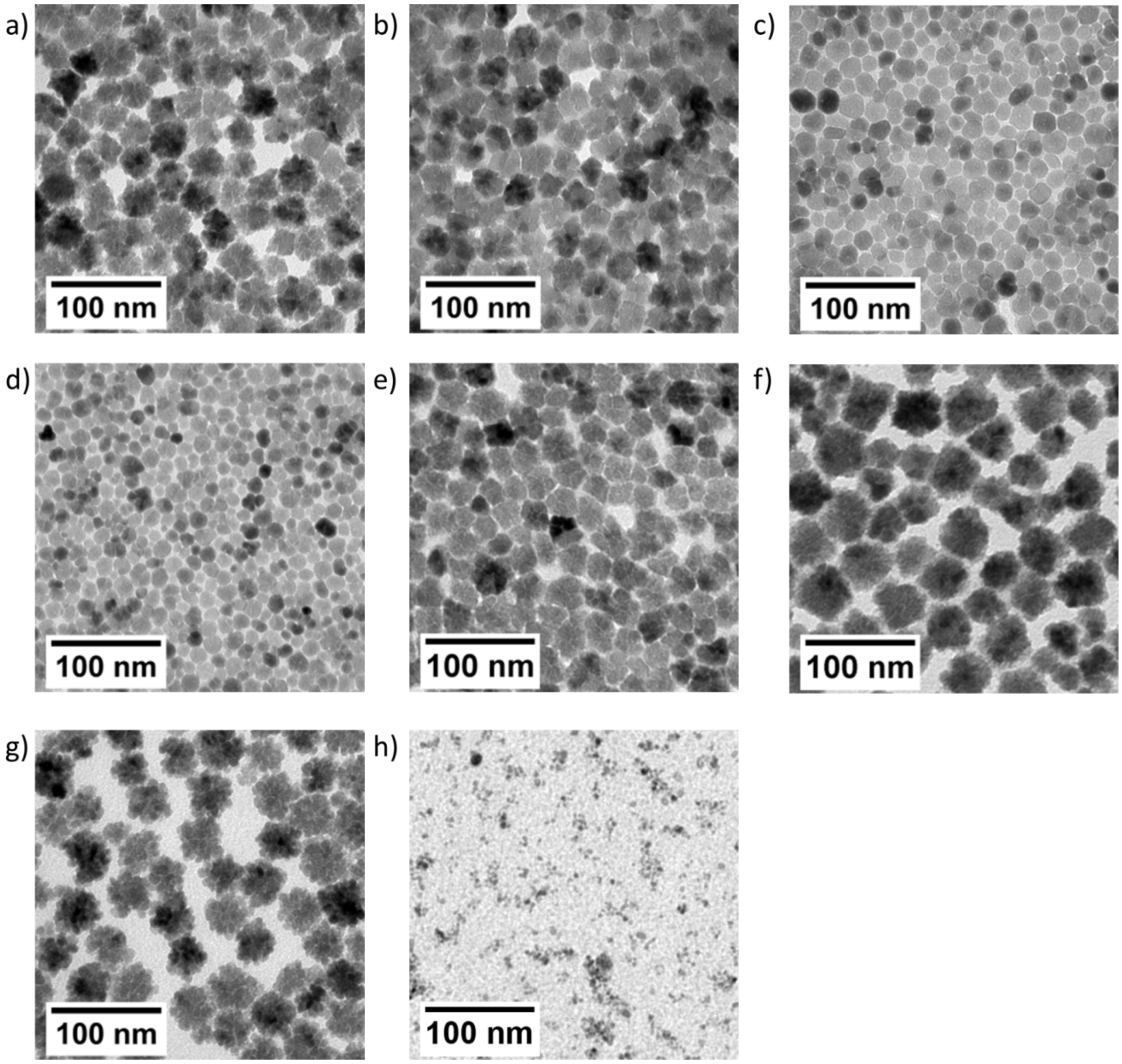

**Figure 1.** TEM micrographs of γ-$Fe_2O_3$ NPs of mono- or multi-core morphology: a) 36ff (32.3±5.0 nm), b) 35ff (29.1±4.4 nm), c) 34ff (18.5±3.2 nm), d) 32ff (14.5±3.4 nm), e) 31ff (27.5±4.2 nm), f) 30ff (46.9±8.5 nm), g) 15ff (36.9±4.8 nm), and h) 17ff (4.3±1.1 nm). TEM sizes in parentheses correspond to the mean outer diameters and respective standard deviations.

Reactants were heated from room temperature (RT) to reflux (approximately 30 min to reach 220 °C) to obtain IONPs. Different quantities of water were injected to the reaction mixture either at RT before heating (*heating-up, HU*) or at reflux (*hot injection, HI*). Solvents used for the syntheses were freshly ordered and preserved from moisture. In the case of the DEG/NMDEA solvent system, water was injected in different amounts, from 100 µL to 2 mL, into 120 mL of the total polyol solvent mixture, representing 0.083 % to 1.67 % volume ratios (v/v) or 5.5 to 110 mmol $H_2O$, *i.e.* stoichiometric (not catalytic) water quantity compared to the total iron(II,III) salts (6 mmol). Lower water amounts did not allow producing IONPs, while larger quantities of water led to ill-defined IONPs. As seen on data (Figure 1 and Table 1), optimal control, in terms of size-distribution and morphology homogeneity, was achieved when mixing a determined amount of water in the solvent system with the precursors before heating (*HU*), as supposedly ascribed to a more homogeneous composition of the starting mixture.

Surprisingly, using anhydrous iron(III) chloride instead of the hexahydrate compound did not allow producing magnetic IONPs, even when traces of water were added before heating (*HU*). This evidences the critical role of water and its necessary presence in the starting iron salt precursors for the positive outcome of the reaction. In the polyol/poly(hydroxy)amine (DEG/NMDEA) synthesis indeed, the solvents act simultaneously as multivalent chelators for iron(II,III) cations, as well as a high boiling temperature medium to achieve a good control over the nucleation and separation from growth of IONPs. The chloride counter-ions of the iron(II,III) salts are thus exchanged by complexing solvent molecules and subsequently by hydroxide anions or by water molecules when adding the NaOH solution. It was described by Caruntu *et al.* that the actual precursors of inorganic polymerization are iron(II,III) hydroxides in which the metallic centers are also chelated by DEG.[11] Therefore, the polyol route is also referred to in literature as a “forced hydrolysis” mechanism. This salt metathesis can be observed by eye when mixing the reactants from the color changes of the organometallic solution turning into a black colloidal suspension of magnetite ($Fe_3O_4$) NPs. Studying the medium of synthesis by $^1$H NMR spectroscopy helped understanding the mechanism of reaction (**Figure S2)**. It was observed that controlled water addition shifts the broad peak attributed to the labile protons of hydroxyl groups in DEG and NMDEA (at respectively 3.43 and 3.07 ppm). Apart from further shifting of this labile protons peak ascribed to pH variation occurring during the synthesis (hydroxyl groups being converted into oxides), the NMR spectrum does not show any evidence of polyol molecules degradation. It is worth noting that the IONPs are, in most cases, covered by a layer of chelating solvent molecules even after several washing steps using a mixture of ethanol and ethylacetate.[14] This affects the colloidal stability of the samples, and their ability to be oxidized by the boiling $Fe(NO_3)_3$ addition.[17] During the oxidation step, the sample color is expected to turn from black to dark red as IONPs are oxidized from magnetite to maghemite. In some cases, especially for nanoflowers, the color of the colloidal solution remained black. This protective layer of DEG and NMDEA at the surface of IONPs was also evidenced by $\zeta$ potential measurement at varying pH (**Figure S3)**. The isoelectric point (IEP) of maghemite is expected at around pH=7, while the IEP of IONPs still covered by a layer of solvent ligands was shifted to about pH=9, which is consistent with the expected pKa value of the tertiary amine in NMDEA. Washing IONPs by a precipitation-redispersion process in aqueous alkaline media revealed to be an efficient means to completely remove the remainder of chelated solvents as evidenced by the shift of the curve of $\zeta$ *vs.* pH after extensive washing and by the ATR-IR spectrum on **Figure S3**.

**Structure and properties of the synthesized IONPs**

In the case of the sole DEG solvent, amounts of water as large as 5 mL, representing 4.2 % of the total solvent volume or 275 mmol $H_2O$ molecules, were injected at reflux temperature with a syringe needle through a septum (*hot injection*, *HI*), generating much smaller IONPs, with diameters typically in the order of 3-5 nm. The fast introduction of a water excess at high temperature immediately generated a sudden nuclei burst, with a solution turning from deep green to black. This hot injection method lead to 'ultra-ultra-small' IONPs (UUSPIOs) as there is a limited quantity of precursors in solution available for crystal growth. The time-scale of reaction was usually much shorter than when synthesizing larger IONPs in DEG/NMDEA. Typically, reactions in sole DEG were completed after only 15-20 min. When letting an aliquot of the reaction mixture at rest over a strong permanent magnet, the supernatant became colorless, evidencing the total conversion of the colored iron(II,III) organometallic precursors into a colloidal magnetite phase.

The final products of all the different batches greatly differed in sizes and shapes according to the solvent composition: only DEG or DEG/NMDEA, or amount and timing of water addition. Smooth spheres as well as more complex structures previously reported as "nanoflowers" were obtained. TEM images enabled to determine both the overall diameters and the individual grain sizes for these multi-core IONPs. These estimates can be compared to other available techniques for particle sizing, namely the fit of the magnetization curve by the Langevin function convolved with a distribution law of diameters, or of the SANS curve by a polydisperse sphere form factor, these two methods being shown in supporting information (**Figures S1 and S4**). Smooth spheres were produced with adjustable sizes from ~4 nm to ~20 nm when the mixture was left at rest during all the reaction (under 'natural convection'), while nanoflowers were obtained when the mixture was continuously homogenized with a stirring shaft, with sizes from ~27 to ~37 nm (**Figure 1**). The size-histograms could be well-fitted using a log-normal distribution law of parameters $\alpha$ and $\beta$ (**Figure 2**). In order to express diameters as $d_0 \pm \sigma$, the mean values $d_0$ along with standard deviations $\sigma$ were calculated according to the following formulas:

$$d_0 = \langle d \rangle = \alpha e^{\beta^2/2} \qquad \text{and} \qquad \sigma^2 = \left\langle (d - d_0)^2 \right\rangle = d_0{}^2 \left( e^{\beta^2} - 1 \right)$$

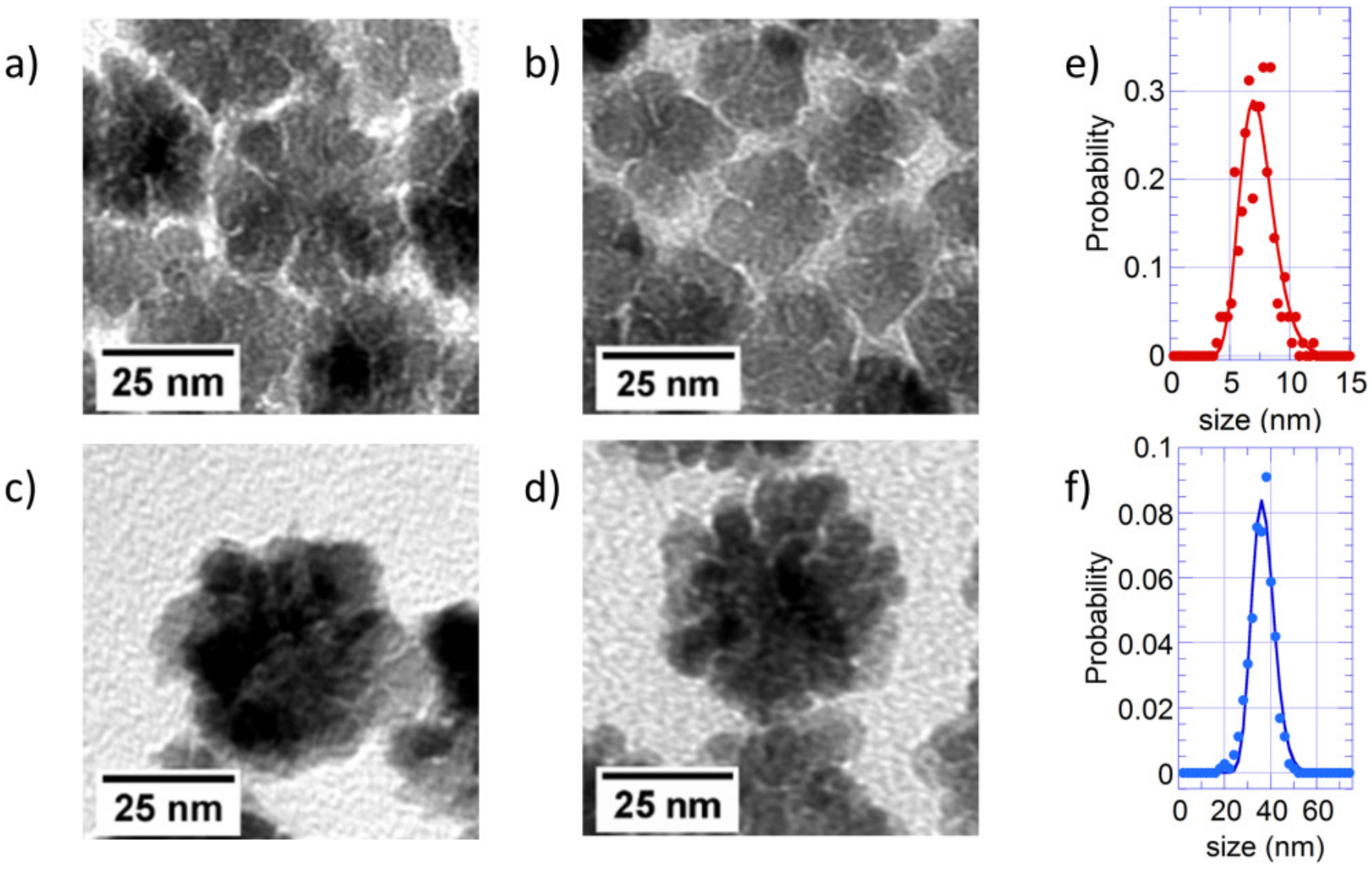

**Figure 2.** High magnification (HM) TEM micrographs of γ-$Fe_2O_3$ nanoflowers: a) 36ff, b) 35ff, c) 30ff, and d) 15ff. Distribution of e) grain sizes of the 15ff batch (7.4±1.4 nm), and f) outer diameter (36.9±4.8 nm), measured on HM-TEM micrographs and log-normal fits.

Size distributions characterized by $\beta$ parameters below 0.2 were considered sufficiently narrow and suitable for further characterization. The condition *N100*$_{HU}$*-5h*, using mechanical stirring, proved to be both optimal and robust for the reproducible synthesis of nanoflowers as evidenced by the similar TEM images obtained for the 35ff and 36ff batches (**Figure 2**). In contrast, similar conditions without stirring (*N100*$_{HU}$*-5h**) led to smooth spherical iron oxide NPs (34ff) of narrow size-distribution 18.5±3.2 nm as determined by TEM analysis, or 18.8±6.5 nm as obtained by fitting the DC magnetization curve (**Figure S4**). This observation highlights the critical role of mixing (diffusion – reaction *vs.* active stirring) as experimental parameter on the resulting IONP morphology.

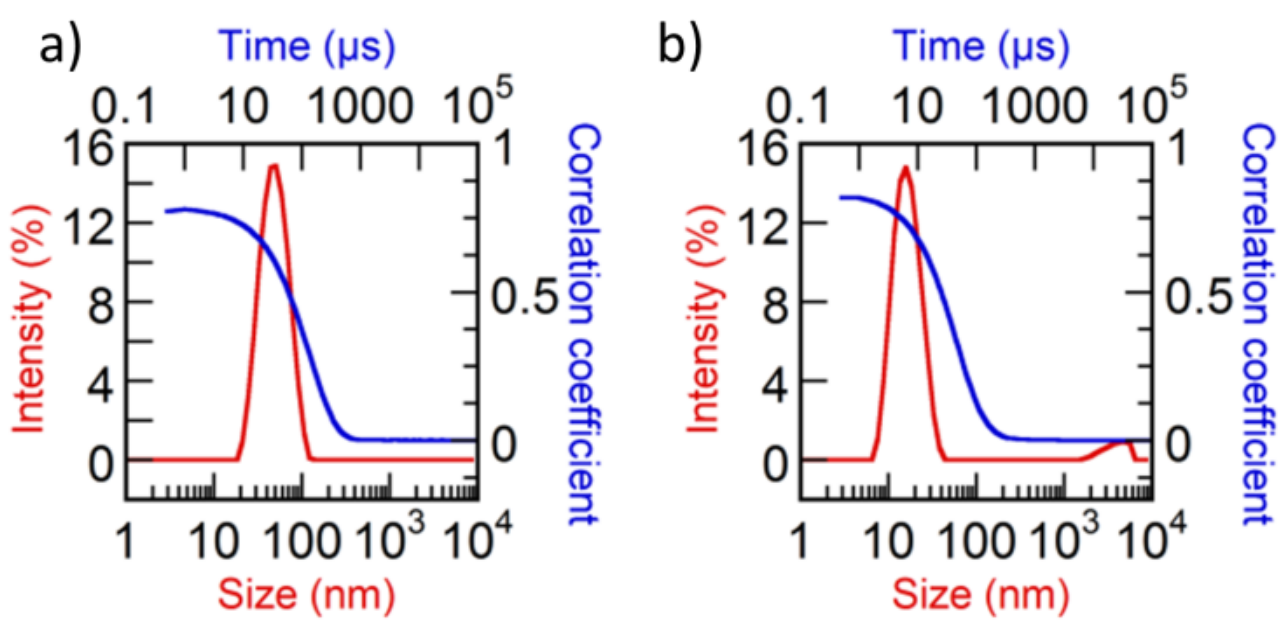

**Figure 3.** DLS correlograms and intensity-averaged distribution of diameters of a) 35ff nanoflowers ($D_h$=36 nm, PDI=0.13), and b) 20ff UUSPIO smooth spheres ($D_h$=16 nm, PDI=0.21). Values in parentheses correspond to the *Z*-average hydrodynamic diameter ($D_h$,) obtained from the 2$^{nd}$ order cumulant fit, and polydispsersity index (PDI). See all DLS plots on **Figure S5**.

On the other hand, the dispersion state of the different IONP batches synthesized was probed by dynamic light scattering (DLS) in a weakly acidic aqueous medium (pH~2). The fit of correlograms by the 2$^{nd}$ order cumulant method provides the *Z*-average hydrodynamic diameter ($D_h$) and polydispersity index (PDI). Typical values of *Z*-average diameters were in the order of 30 nm, ranging from 16 to 55 nm when considering all synthesized batches (**Figure S5**). For instance, the *Z*-average hydrodynamic diameter of 35ff nanoflowers was measured at $D_h$=36 nm (PDI=0.13), while 20ff UUSPIO smooth spheres were sized at $D_h$=16 nm (PDI=0.21) (**Figure 3**). Hydrodynamic diameters estimated by DLS are always found larger than diameters determined by TEM due to the dried state of the TEM sample. The DLS sizes however represent better the real dispersion state of NPs.

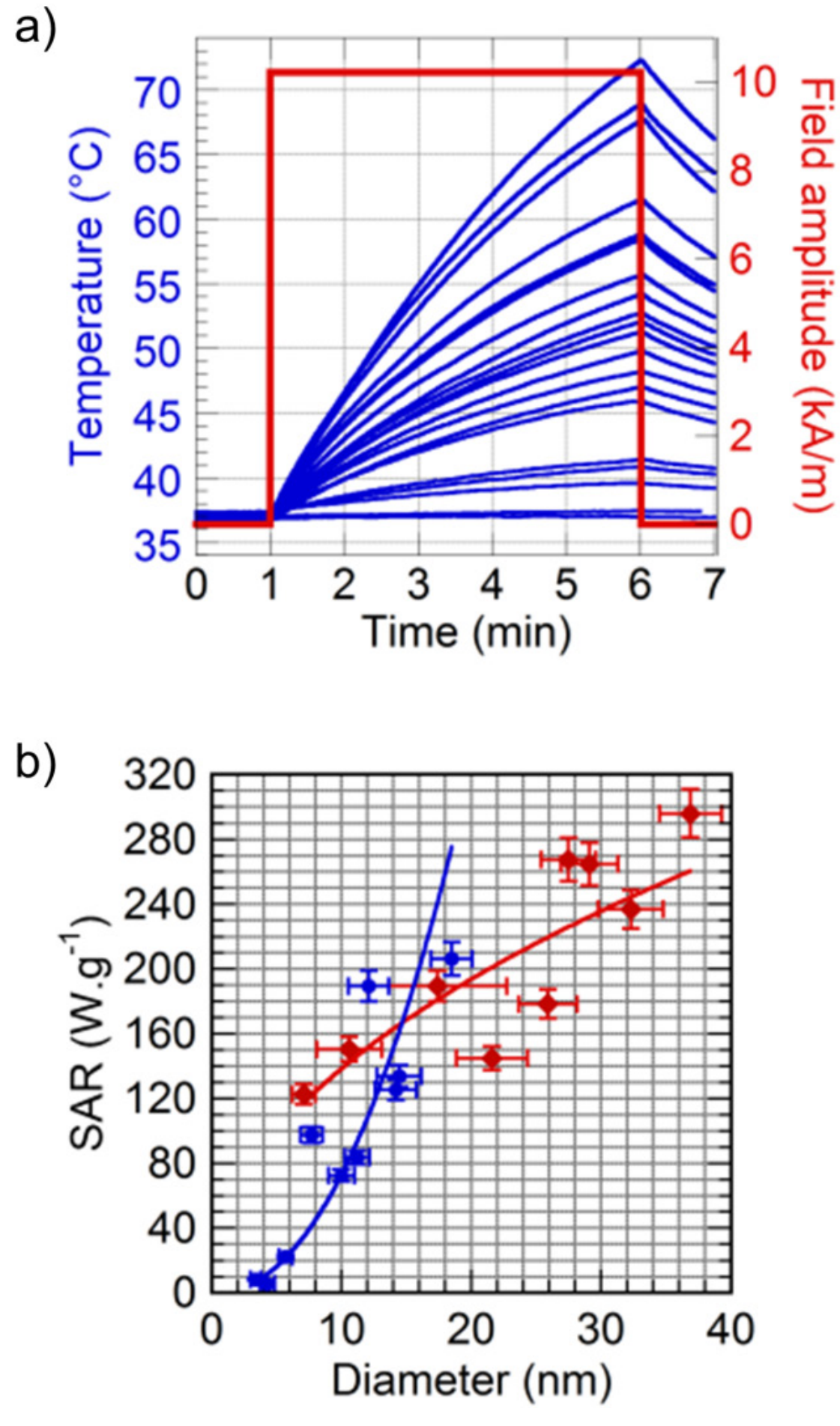

**Figure 4.** a) Temperature profiles *vs.* time of different samples of iron oxide concentration set at 3 g·L$^{-1}$, under application of an AMF at $H_{app}$=10.2 kA·m$^{-1}$ and $f$=755 kHz. The full list of samples and corresponding initial slopes (within first 5 s) is given, on **Table S1**; b) Deduced *SAR* *vs.* mean diameter (determined by TEM) for smooth nanospheres (blue) and nanoflowers (red). Solid lines are power law fits of exponents 2.2 and 0.48, respectively.

The efficiency of the different NP batches for magnetic hyperthermia was evaluated by applying an AMF for 5 min and by recording the temperature rise of the solvent, with iron oxide concentrations set to 3 g·L$^{-1}$. While no significant heating under AMF was obtained for a few samples like the 17ff UUSPIOs with the smallest size, a temperature rise from 37 °C to 70 °C was recorded for the largest dispersible nanoflowers 15ff (**Figure 4**). Therapeutic hyperthermia requires that IONPs heat cancerous tissues up to 43–44 °C to deposit a suitable "thermal dose", usually by IONP intratumoral injection.[28] This temperature may potentially be reached *in vivo* with the best heating samples after few minutes, even at a concentration as low as the one used in our study (3 $g{\cdot}L^{-1}$), assuming that their heating properties are preserved in physiological intracellular conditions.[29] The heating properties of our sample library were thus quantified using the specific absorption rate (*SAR*) determined experimentally using the commonly used formula:

$$SAR = \left(\Delta T/\Delta t\right)_{t\to 0} C_P/m$$

where $(\Delta T/\Delta t)_{t\to 0}$ is the temperature raise slope at early times of AMF application (first 5 s) to simulate adiabatic conditions,[30] $m$ is the mass of nanoparticles in 1 mL of suspension and $C_P$ is approximated by the specific heat of pure water.[2]

We have used the *SAR* to evaluate the heating properties of our nanoparticles instead of the intrinsic loss power (*ILP*)[31], because the *SAR* variation with field intensity can depart significantly from a quadratic law. The plot of *SAR* at given field amplitude ($H_{app}$=10.2 $kA{\cdot}m^{-1}$) and frequency ($f$=755 kHz) *vs.* diameter measured by TEM clearly evidences a correlation between the *SAR* and IONP outer diameter measured by TEM. (**Figure 4b**) Nanospheres experimentally follow a quadratic law with diameter while nanoflowers follow a lower exponent (nearly square-root). This is qualitatively in agreement with the most advanced models on the optimal size of magnetic NPs for MFH at given values of their other physical properties (specific magnetization and magnetic anisotropy).[32] Results compiled in Table S1 evidence that all nanoflowers and smooth spheres larger than 12-14 nm are efficient nanoheaters, while UUSPIOs smaller than 5-6 nm do not generate sufficient heat but are likely useable as positive ($T_1$-weighted) MRI contrast agents, as shown later in this article. Smooth spheres of intermediate diameters (10-14 nm) are ideal to be used both as nanoheaters for MFH and as negative ($T_2$- weighted) MRI contrast agents (*vide infra*), once coated with appropriate ligands, while larger ones might be more difficult to stabilise in physiological media.

An in-depth characterization of magnetic heating properties was carried out on selected IONP samples with interesting morphologies and core size in 20-40 nm range as evidenced by TEM, with outstanding heating efficiency ($SAR$>200 W·g$^{-1}$ at $H_{app}$=10.2 kA·m$^{-1}$ and $f$=755 kHz). Large smooth spheres (34ff) and nanoflowers of different grain size and increasing outer diameters (31ff, 35ff, and 15ff) were thus selected to be further examined with an in-house developed pick-up coil set-up, allowing $SAR$ measurement on a broad range of AMF frequencies and amplitudes.[33] AC magnetization curves of NPs are plotted on **Figure 5a** *versus* amplitude $H_{app}$ up to 21 kA.m$^{-1}$ at fixed frequency ($f$=1030 kHz). Similar curves at different frequencies are provided as supporting information accompanying this manuscript (**Figures S6, S7, S8 and S9**).

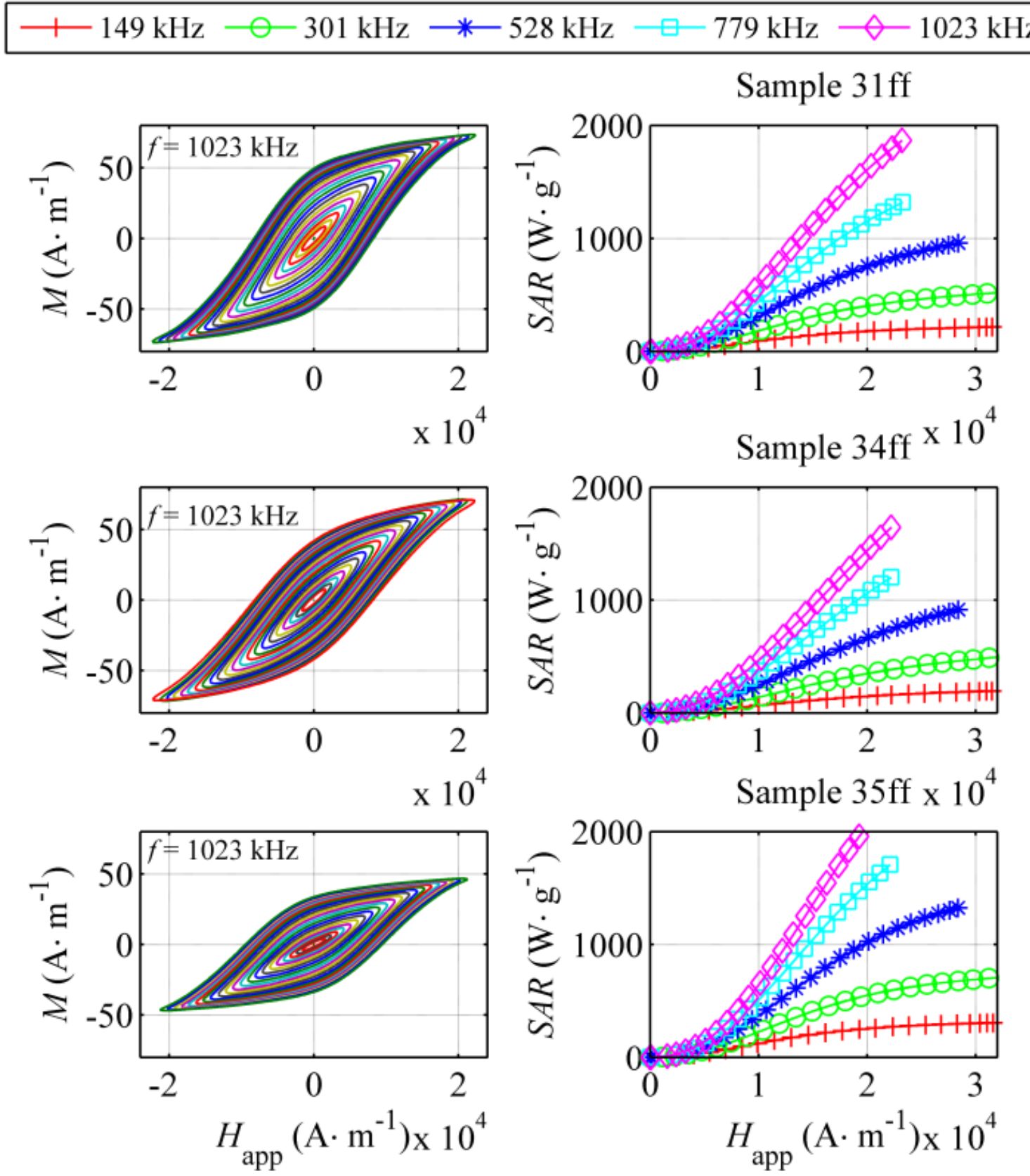

**Figure 5.** a) AC hysteresis loops of samples 31ff, 34ff and 35ff at 1023 kHz fixed frequency (left); b) *SAR* of samples 31ff, 34ff and 35ff *versus* field amplitude at different frequencies (right).

Such AC hysteresis loops reflect that magnetic moments of IONPs under AMF excitation oscillate with a phase lag relatively to the magnetic field, leading to a partial conversion of the radiofrequency magnetic energy into heat that then dissipates in the surrounding aqueous medium. Larger AC hysteresis loops areas are obtained at higher amplitudes $H_{app}$ and higher frequencies, while stronger magnetic anisotropy of the materials tends to change the shapes of the curves from sigmoidal to more square-like shapes. Carrey *et al.* interpreted such dynamic hysteresis loops by a two-level Stoner-Wohlfarth model instead of the classical linear response theory involving Néel and Brown relaxation times of the magnetic moments.[32, 34, 35] The larger the surface of the hysteresis loops, the larger the energy dissipated by the IONPs per AMF cycle. These AC magnetization curves can be compared to the calorimetric experiments by multiplying the surface of the hysteresis loop with the frequency and by dividing by iron oxide concentration to get the *SAR*. The resulting values are then plotted as a function of frequency and amplitude, as shown in **Figure 5b**.

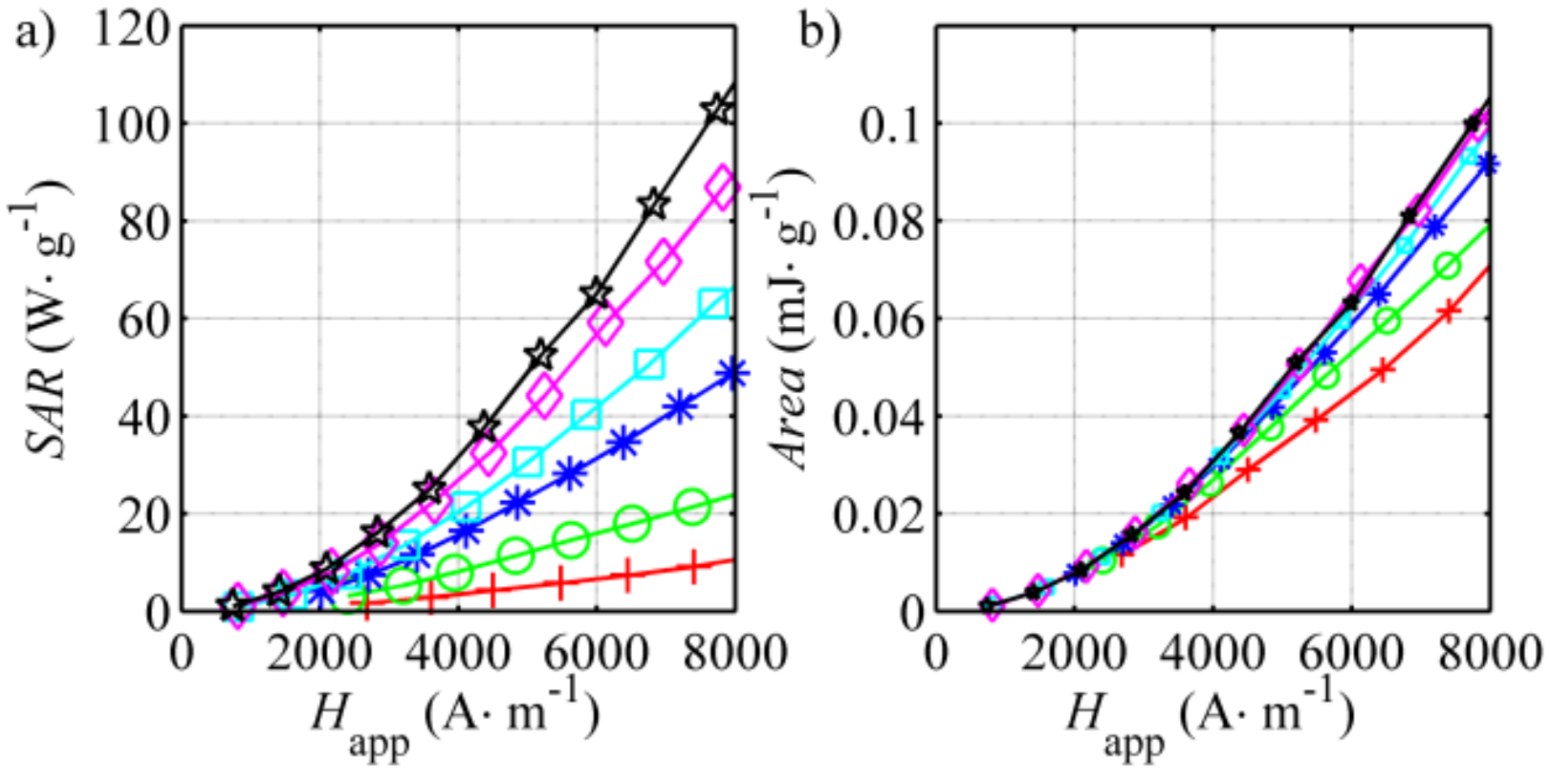

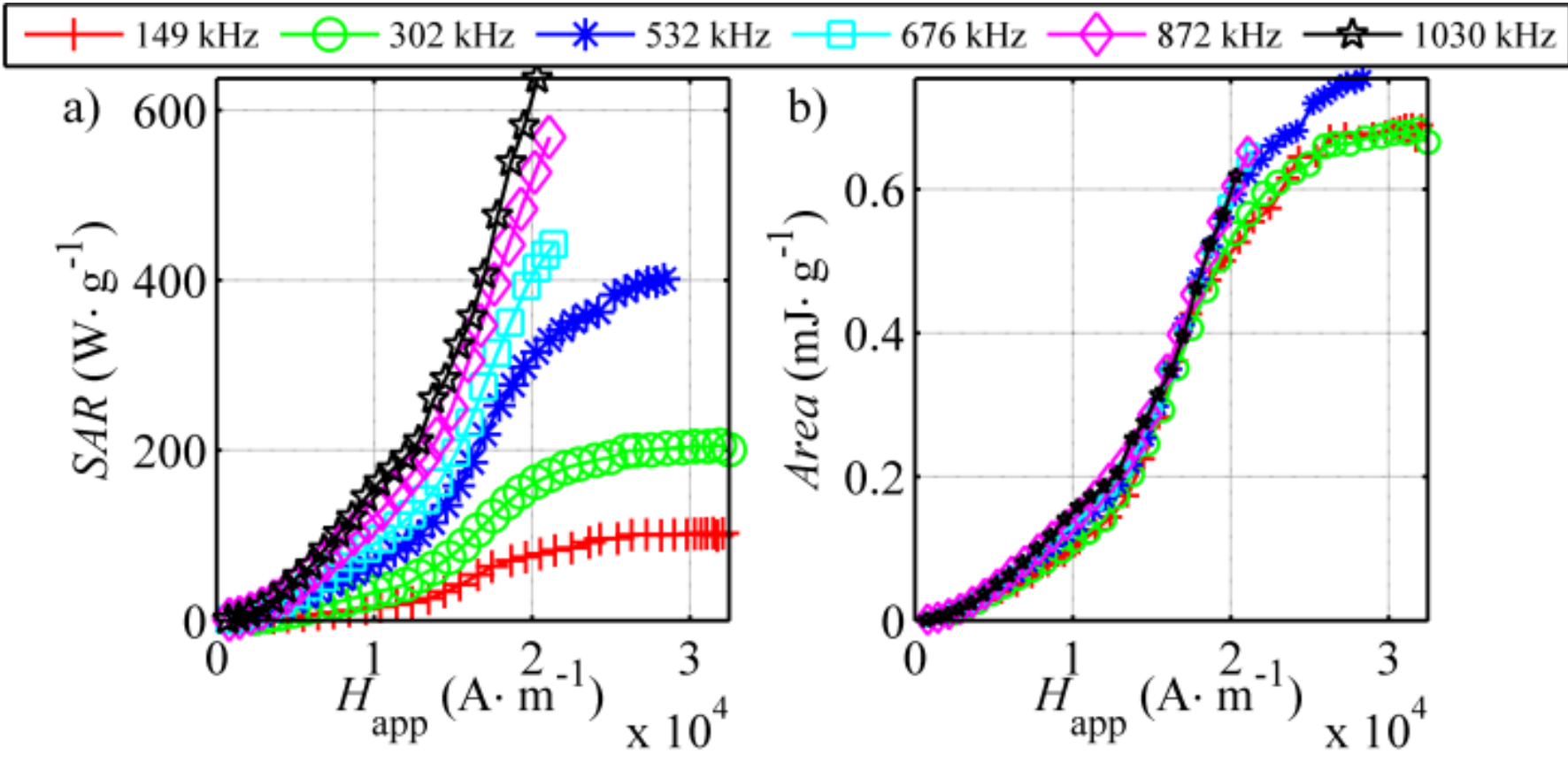

**Figure 6.** a) *SAR* and b) hysteresis loop area *A* *vs.* applied intensity $H_{app}$ of the oxidized 15ff nanoflower batch. As can be appreciated on the zoomed curves at low $H_{app}$ (two upper graphs), the 15ff sample behaves as a superparamagnetic materials at low field (*SAR* varies as the square of $H_{app}$). At higher field (bottom graphs), it rather shows ferromagnetic behavior with a hysteresis area that first rises above threshold anisotropy field $H_{an}$ and then reaches a plateau.

*SAR* values as high as 2000 $W{\cdot}g^{-1}$ were obtained at a frequency of 1023 kHz and amplitude of 20 $kA{\cdot}m^{-1}$, although out-passing by a factor 4 the upper limit of $5{\times}10^{9}$ $A{\cdot}m^{-1}{\cdot}s^{-1}$ of the $f{\cdot}H_{app}$ product recommended for human treatment by MFH.[36] Whereas the *ILP* parameter was previously introduced in the literature by dividing the *SAR* with frequency and the square of $H_{app}$,[1] the assumption that *SAR* is a quadratic function only works at low field amplitude (**Figure 6a**). We have therefore preferred to report the ratio (in $J{\cdot}g^{-1}$) of the *SAR* and frequency, which represents the area *A* of a hysteresis loop (*i.e.* thermal losses per AMF cycle). This enables to directly compare the heating properties of the magnetic IONPs synthesized in this work to different ones reported in literature, although measured under various conditions and setups. As reported by Carrey *et al.*,[32] the plot of hysteresis surface area *vs.* $H_{app}$ expected for ferromagnetic NPs exhibits a sigmoidal shape: it starts with a slow increase as long as the AMF amplitude is lower than the anisotropy field $H_{an}$ of the nanomaterials, then increases rapidly (*i.e.* with an exponent larger than two, value expected for superparamagnetic NPs), and finally plateaus in the high-field limit. The plot of $A$ *vs.* $H_{app}$ for the 15ff sample perfectly fits this description (**Figure 6b**), with a threshold anisotropy field $H_{an}$~$10^4$ A·m$^{-1}$. The plateau value around 0.7 mJ·g$^{-1}$ for the oxidized 15ff sample is not particularly outstanding as a maximal value of 1.8 mJ·g$^{-1}$ was previously reported for IONPs obtained by coprecipitation followed by hydrothermal treatment.[37] The oxidized 15ff sample however illustrates the complex magnetic behaviour of nanoflowers reflected in the dependence of their hysteresis loss area $A$ with the amplitude $H_{app}$ of the AMF: For the six probed frequencies, the plots of the hysteresis area $A$ *vs.* $H_{app}$ collapse almost perfectly on a unique master curve. The field dependence remains quadratic up to a threshold $H_{app}$ ascribed to the anisotropy field $H_{an}$ of the multi-core structure, characteristic of collective dynamics of sintered grains as in a multiple-domain magnet. For any applied AMF strength below $H_{an}$, nanoflowers exhibit pure superparamagnetic response as evidenced by the quadratic variation of their *SAR vs.* $H_{app}$ plot, each of their magnetic mono-domains being excited individually by the AMF. Other IONP batches of lower outer sizes (below 30 nm) exhibit even superior plateau values of the hysteresis area per cycle *vs*. $H_{app}$ (which exact value slightly varies with frequency), from ~1.6 mJ·g$^{-1}$ for 31ff nanoflowers (**Figure S10)** and 34ff smooth nanospheres (**Figure S11)** to ~2.5 mJ·g$^{-1}$ for 35ff nanoflowers (**Figure S12**) or a bit lower (2 mJ·g$^{-1}$) at the lowest frequency of 149 kHz. To the best of our knowledge, these are the highest hysteresis area values reported so far for synthetic magnetic IONPs, excluding the case of needle-like submicron $\gamma$-$Fe_2O_3$ particles commercialized for magnetic recording applications, that can reach hysteresis areas up to 8 mJ·g$^{-1}$, but necessitate to apply an AMF stronger than a thrice higher threshold field, $H_{an}$~30 kA·m$^{-1}$.[38]

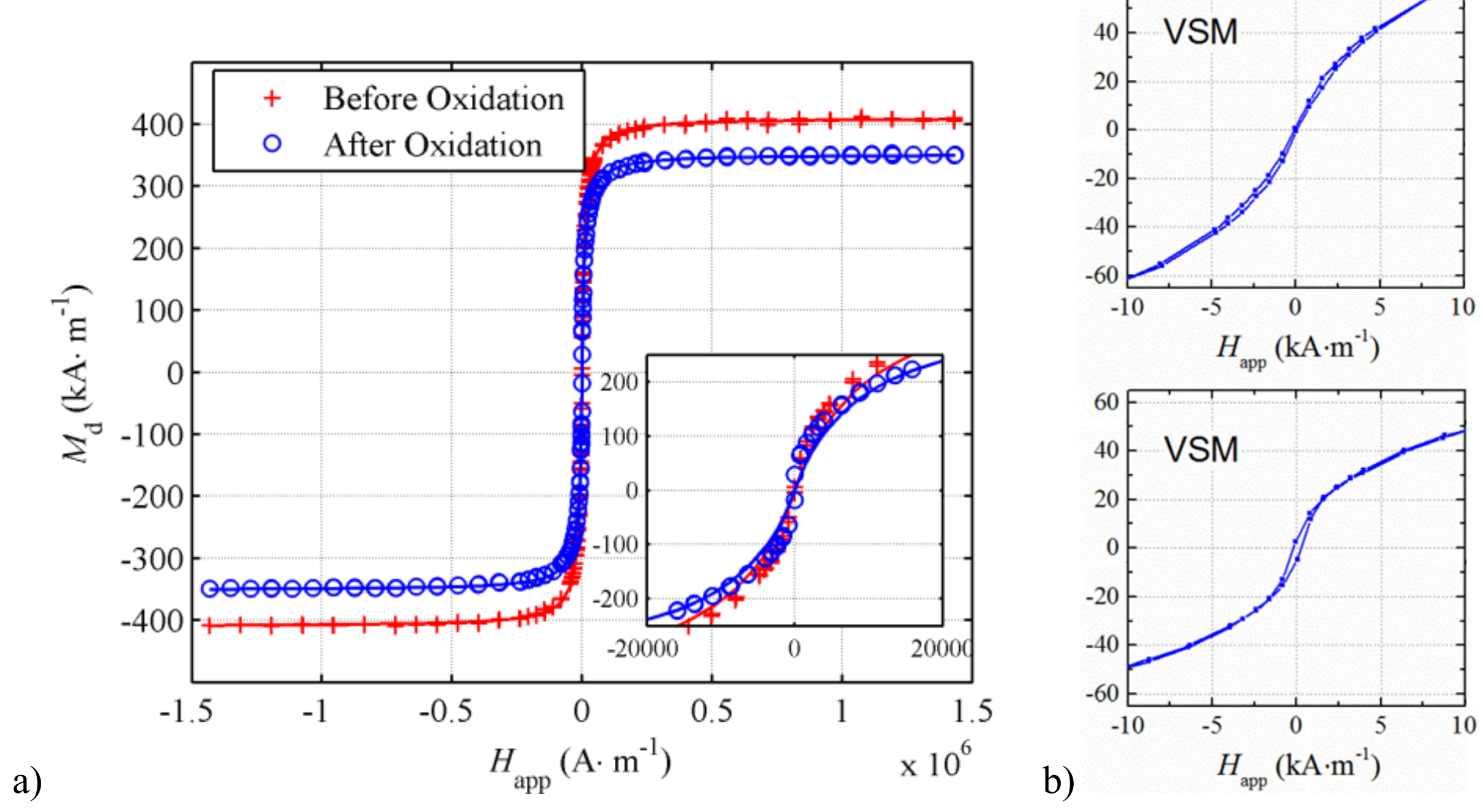

**Figure 7.** a) Static magnetization curve of 15ff nanoflowers measured by VSM at room temperature in water before and after oxidation (red and blue curves, respectively). In a narrow range $-1.6<H_{app}<1.6$ kA·m$^{-1}$ only, the curves can be approximated by $M_d=\chi\cdot H_{app}$, the susceptibility reaching a value as high as $\chi=59$ in both cases; b) Enlargements of the VSM curves in the low field region ($-10<H_{app}<10$ kA·m$^{-1}$) of 15ff nanoflowers dispersed in an agar gel (upper curve) or in glycerol (lower curve), showing that the coercive field ($H_c$) and remnant magnetization ($M_r$) – if any – lay near the limit of sensitivity of the field probe *i.e.* ~500 A·m$^{-1}$.

To get an insight into the peculiar structure-property relationships and complex magnetic behavior of nanoflowers, the DC magnetization curves of 15ff IONPs (with or without the oxidation step) were also measured with a lab-made VSM setup (**Figure 7**). In both cases, the IONPs were dispersed in water acidified with dilute $HNO_3$ (pH~2.5). The magnetization curve of both samples exhibits null coercive field ($H_c$=0) and zero remanence ($M_r$=0), which corresponds to pure superparamagnetic behavior. This evidences the good dispersion state of these IONPs in water, as a remnant magnetization would have been expected in case of an aggregated sample. The saturation magnetization of the non-oxidized sample obtained after synthesis and washings is 350 $kA\cdot m^{-1}$, below the value of bulk magnetite ($Fe_3O_4$) (500 $kA\cdot m^{-1}$), as ascribed to spin-canting defects at the NP surface,[39] or to partial oxidation already starting during the purification steps, as no particular precautions were taken to prevent it. In contrast, total oxidation of sample 15ff obtained by heating with iron(III) nitrate, led to a magnetization at saturation of 300 $kA\cdot m^{-1}$, which is again below the expected value of 400 kA m-1 for bulk maghemite ($\gamma$-$Fe_2O_3$) but nevertheless quite a satisfying value for IONPs. Therefore, all the IONP batches were oxidized intentionally to control the magnetic phase of the IONPs, despite lowering the magnetic saturation and, presumably, their heating efficiency for MFH. All samples reported in this study exhibit similar saturation magnetization (**Figure S4**).

A second useful information provided by VSM magnetometry can be obtained by fitting the DC magnetization curves by the Langevin function characteristic of superparamagnetism, convolved by a log-normal distribution of diameters to take into account size-dispersity.[21] The resulting magnetic domain diameters lay below the outer diameter measured by TEM for nanoflowers: 25.1±12.0 nm for 35ff, 21.9±10.6 nm for 31ff, while almost identical to TEM (within experimental uncertainty) for 34ff smooth nanospheres: 18.8±6.5 nm.

Sensitive magnetometry performed on the commercial MPMS system on the oxidized and non-oxidized 15ff nanoflowers (**Figure S13**) led to ZFC and FC magnetization curves *vs.* temperature both lower for the non-oxidized as compared to the oxidized 15ff NPs. Such non-classical ZFC-FC curve profile has been already reported for large (18 and 22 nm) $Fe_3O_4$ NPs synthesized by iron(III) oleate thermal decomposition,[40] and was partially explained by the so-called Verwey or charge ordering transition,[40, 41] when such IONPs undergo a slight crystallographic distortion from cubic, electrically conducting, to inverse spinel, electrically insulating, structure, this change of crystalline structure also impacting the magnetic properties. A suitable method to estimate the blocking temperature $T_B$, defined at the transition from the ferrimagnetic state to the superparamagnetic regime, consists in plotting the derivative of the $M_{FC}$-$M_{ZFC}$ curve difference with respect to temperature.[22] For the two different batches (non-oxidized and oxidized 15ff), the plot exhibits three maxima (**Figure S13**). The peak near 90 K is ascribed to the Verwey transition of magnetite, yet it is not clear why it also appears on the oxidized sample. The two other peaks correspond to characteristic temperatures, respectively near 200 and 300 K. It is rather uncommon for a sample to exhibit two values of blocking temperatures. One hypothesis is that $T_{B1}$≈200 K is ascribed to individual magnetic domains of diameter 7.4±1.4 nm and $T_{B2}$≈300 K to the whole magnetic multi-core structure of outer diameter 36.9±4.8 nm. The complete interpretation of these data, *e.g.* in term of superspin glass transition, would necessitate complementary AC susceptometry experiments *vs.* temperature, to study the slow relaxation dynamics of the frustrated spins in multi-core nanoflowers as reported by Kostopoulou *et al.*,[10] but it is beyond the scope of this article.

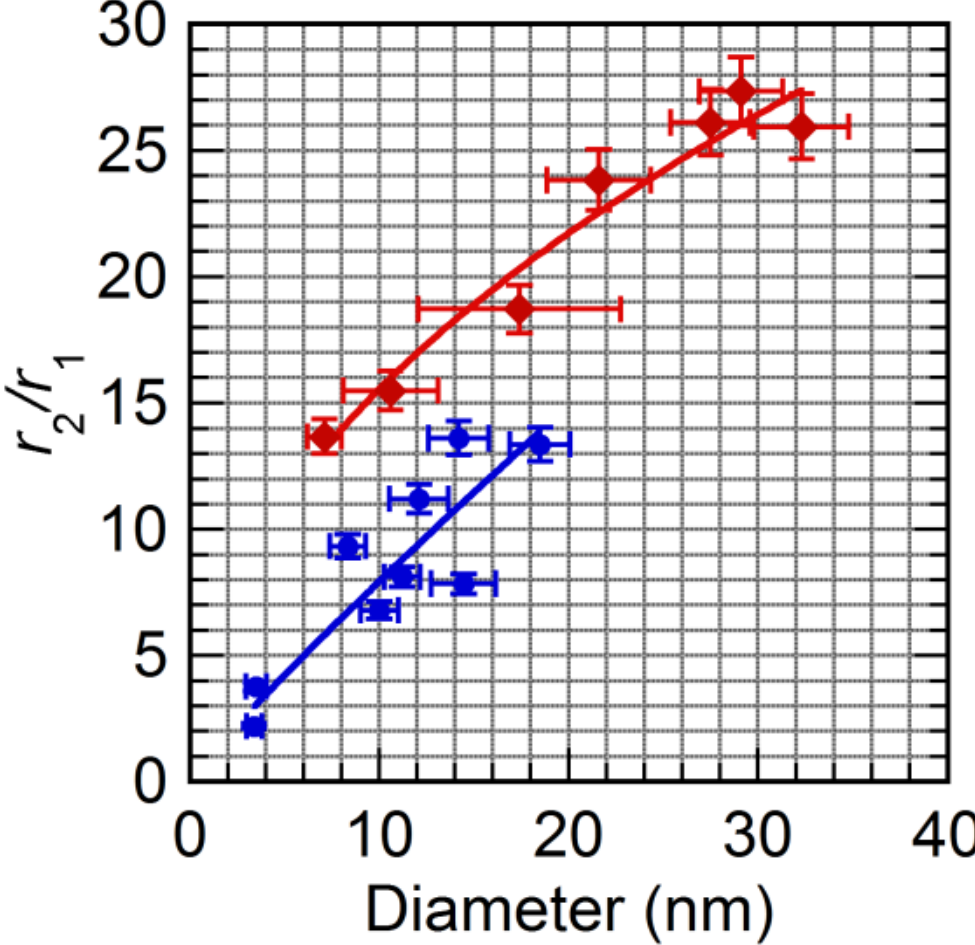

**Figure 8.** Ratio of transverse ($r_2$) to longitudinal ($r_1$) relaxivities at 1.41 T/60MHz (37°C) *vs.* average TEM diameters of IONPs, for nanosphere (blue) and nanoflower (red) samples. Solid lines are power law fits of exponents respectively 0.92 and 0.48.

In order to assess the efficiency of the different batches synthesized by the polyol route to relax nuclear spins of water protons, their transverse ($r_2$) and longitudinal ($r_1$) relaxivities were measured at physiological temperature (37 °C) with a 60 MHz relaxometer based on a 1.41 Tesla magnet (*i.e.* close to the 1.5 Tesla magnetic field of most clinical MRI machines used in hospitals) (**Figure 8**). Practically, the longitudinal ($T_1$) and transverse ($T_2$) relaxation times of water protons were measured with an inversion-recovery and a CPMG sequence, respectively at decreasing equivalent [Fe] concentrations starting from 6 mM, in acidified water to maintain colloidal stability. Relaxivities are obtained from the slope of the linear variation with [Fe] of the longitudinal (respectively transverse) decay rate of water proton spins, according to:[19]

$$1/T_{i=1 \text{ or } 2} = r_i \times [\mathrm{Fe}] + (1/T_i)_{\mathrm{water}}$$

where $(1/T_i)_{\mathrm{water}}$ represents diamagnetic contribution of water.

In the “outer sphere” model of MRI contrast agents introduced by Ayant and Freed for paramagnetic agents, and adapted by Gillis *et al.* to superparamagnetic IONPs,[42] the increase relaxation rate $1/T_2$ compared to pure water originates from fluctuating dipolar interactions between nuclear spins of water protons and the electronic magnetic moment of IONPs. For a limited range of diameters called "motional averaging regime", the superparamagnetic particle can be considered immobile during the echo time (TE) of the sequence compared to random trajectories of water molecules diffusing all around the magnetic sphere. In this case, Vuong *et al.* have shown that $r_2$ follows a universal scaling law that is quadratic both with the magnetization and with the radius of the "outer sphere", defined as the minimum approach distance between $H_2O$ molecules and the IONP center.[43] For the IONP batches prepared in this study, the quadratic law is perfectly observed for the smooth sphere NPs (**Figure S14**), validating the proton diffusive model. This brings additional evidence that no organic layer remains on their surface after the washing steps. If not, water protons could not reach IONP surface and the quadratic law would not be observed. In the case of nanoflowers, the variation of $r_2$ *vs.* size appears erratic, presumably because of their rough geometry and very high specific area mentioned before, that can be up to 60% larger than the geometrical area of a smooth sphere (see case of ff35 batch on **Figure S1**). Proton relaxivity may arise by a combination of "outer sphere" and "inner sphere" mechanisms, meaning that water molecules can be transiently adsorbed in the porosity of nanoflowers, relaxation dynamics can therefore not be modelled by a single translational diffusion constant of water molecules around the particle.

The ratio of relaxivities $r_2/r_1$ is commonly used to determine whether NPs are most suitable as $T_1$ (positive) or $T_2$ (negative) MRI contrast agents. With $r_2/r_1$ ratios larger than 5, most of the IONPs synthesized here are suitable as negative contrast agents for $T_2$ weighted MRI applications, such as commercial medical products Resovist®, Feridex®, Cliavist® or Clariscan®.[44] With much smaller $r_2/r_1$ ratios, UUSPIOs of just a few nm diameters synthesized in pure DEG would rather be perfectly suitable as $T_1$-type, positive contrast agents. Such UUSPIO-based $T_1$-type contrast agents not yet commercially available would however be an excellent alternative to gadolinium chelates currently used and suspected of toxicity.[45] Free gadolinium ions (*i.e.* not complexed by very strong ligands) are indeed nephrotoxic and neurotoxic, and undesirable side effects may arise in patients with renal insufficiency. Iron oxides having been proven perfectly safe in clinical use, these UUSPIOs are likely to find applications as alternative $T_1$ contrast agents in clinical MRI in a near future.

## CONCLUSIONS

We have presented herein a comprehensive study of IONPs synthesis and characterization. IONPs were synthesized by the forced hydrolysis of iron(II, III) chlorides either in pure DEG polyol or in a mixture with NMDEA poly(hydroxy)amine. We have evidenced the importance of controlling the amount and timing of water addition for a successful synthesis and proper control over IONP morphology. A library of samples was obtained, ranging from “ultra-ultra-small” (~3 nm) UUSPIOs obtained by water “hot injection” in DEG at 220°C and fast growth (20 min), to large IONPs (up to 37 nm) obtained by a longer “heating up” protocol in a DEG/NMDEA 1:1 mixture. Depending on reaction conditions (natural mixing *vs.* mechanical stirring), either smooth sphere or nanoflower morphologies were obtained. The structural and magnetic properties of these nanoparticles were extensively studied. They all exhibit superparamagnetic behavior characterized by a reversible magnetization curve in static magnetic field, with a strong saturation magnetization above $3\times10^5$ $A\cdot m^{-1}$. On the physical point of view, the specific absorption rate (*SAR*) was first tested at given AMF conditions (755 kHz, 10.2 $kA\cdot m^{-1}$) and iron oxide concentration (3 $g\cdot L^{-1}$). These parameters were also varied over a broad range of field strength and frequencies with the use of an AC magnetometer that is much faster than calorimetric experiments for estimating the *SAR*. The efficiency of most of synthesized batches as MRI contrast agents were also evaluated by proton relaxometry measurements. Several scaling laws were derived for the *SAR* and the relaxivity ratio $r_2/r_1$, both being estimated at physiological body temperature (37 °C). At given AMF condition, the *SAR* exhibits quadratic variation with diameter for smooth nanospheres and slower variations (*viz.* square root) for nanoflowers. Transverse relaxivity $r_2$ exhibits also quadratic variation with diameter for smooth nanospheres, in line with the "motional averaging regime" of the "outer sphere" model of MRI contrast agents. The $r_2/r_1$ ratios, calculated to distinguish IONPs better suited as $T_1$ or $T_2$ MRI contrast agents, varies linearly on diameter for nanospheres, and with a lower exponent (*viz.* square root) for nanoflowers. In the case of the smallest diameter IONPs synthesized (USPIOs), their longitudinal relaxivity $r_1$ at 1.41 T associated with moderate $r_2/r_1$ ratio make them alternatives to gadolinium chelates as positive MRI contrast agents, with lower risk of side effects on patients.

The AC hysteresis loops measured at varying AMF frequency and intensity brought more information on the magnetic hyperthermia mechanisms. In the case of nanoflowers and large nanospheres, the *SAR vs.* field intensity curve shows an inflexion point between low and high fields. Below this threshold field, they present perfect superparamagnetic behavior, whereas above this field they behave more like ferromagnets. The threshold field can be associated with an anisotropy field as was done in the two-level Stoner-Wohlfarth model developed by Carrey *et al.* for blocked magnetic moments.[32]

Besides these obvious applications as nanoheaters in magnetic fluid hyperthermia and as MRI contrast agents, other uses can be envisioned. These very large diameter IONPs yet forming stable colloidal suspensions exhibit extremely large magnetic susceptibility in DC magnetic field ($\chi$ up to 100), which can be a requested property for delicate experimental setups based on weak DC magnetic fields such as the "magnetic tweezers" to manipulate living tissues or embryos,[46] or micromechanical experiments to assess the flexural rigidity of magnetic wires.[47]

Another application could also be magnetic particle imaging (MPI) that requires IONPs optimally in the 20-25 nm diameter range so that magnetization saturates at low field strengths.[48] Such versatility of sizes, morphologies and thus of physical properties was achieved by playing only on the nature of polyols as well as the amount and way of introducing water in the reaction vessel ("hot injection" *vs.* "heating up"), in solvent reflux conditions. In brief, robust, gram-scale and easily reproducible synthesis protocols were described to prepare from ultra-ultra-small superparamagnetic cores to very large size magnetic smooth nanospheres and nanoflowers, the latter offering among the highest magnetic heating properties reported so far for synthetic iron oxide nanoparticles.

## ASSOCIATED CONTENT

**Supporting Information**. The Supporting Information is available free of charge on the ACS Publications website at DOI: 10.1021/acs.inorgchem.7b00956.

SANS curves, NMR analyses, Zeta potential, ATR-IR spectrum, DC magnetization curves, DLS curves, Summary of characteristics of the sample library, AC hysteresis cycles, SAR and hysteresis area vs. applied magnetic field at constant radiofrequency, ZFC-FC curves by MPMS magnetometry, longitudinal and transverse relaxivities (PDF).

## AUTHOR INFORMATION

### Corresponding Author

* eneko.garayo@ehu.eus and olivier.sandre@enscbp.fr

### Author Contributions

The manuscript was written through contributions of all authors. All authors have given approval to the final version of the manuscript.

### Funding Sources

G.H.'s doctoral fellowship was funded by the 2014 call of the Department of Science and Technology of the University of Bordeaux (APUB1–ST2014). A. K.'s internship was supported by the US-France-Belgium iREU Site in Translational Chemistry funded by the National Science Foundation (Grant No. CHE 1560390) headed by Pr Randall J. Duran. Basque Government (Grant No. IT-1005-16) and Agence Nationale de la Recherche (Grant ANR-13-BS08-0017 MagnetoChemoBlast) are also acknowledged for financial support. This article is based upon work from COST Action RADIOMAG (TD1402), supported by COST (European Cooperation in Science and Technology).

## ACKNOWLEDGMENT

Laboratoire Léon Brillouin (CEA-Saclay) also is acknowledged for giving access to G.H. to the PACE spectrometer during the 2015 annual training on neutron scattering techniques (Fan du LLB/Orphée). TEM images were taken at the Bordeaux Imaging Center (BIC) of the University of Bordeaux with the acknowledged help of Sabrina Lacomme and Etienne Gontier on equipment funded by France Life Imaging. Authors also want to thank SGIker (UPV/EHU) for the technical and human support as well as the VSM and ZFC-FC measurements.

## ABBREVIATIONS

AMF, alternating magnetic field; DEG, diethylene glycol; DLS, dynamic light scattering; ATR-IR, attenuated total reflection infrared spectroscopy; *ILP*, intrinsic loss power; IONP, iron oxide nanoparticle; IEP, isoelectric point; MFH, magnetic fluid hyperthermia; MPMS, magnetic property measurement system; NMDEA, *N*-methyl diethanolamine; NP, nanoparticle; OD, optical density (or absorbance); SANS, small angle neutron scattering; *SAR*, specific absorption rate; UUSPIO, ultra-ultra-small superparamagnetic iron oxide.

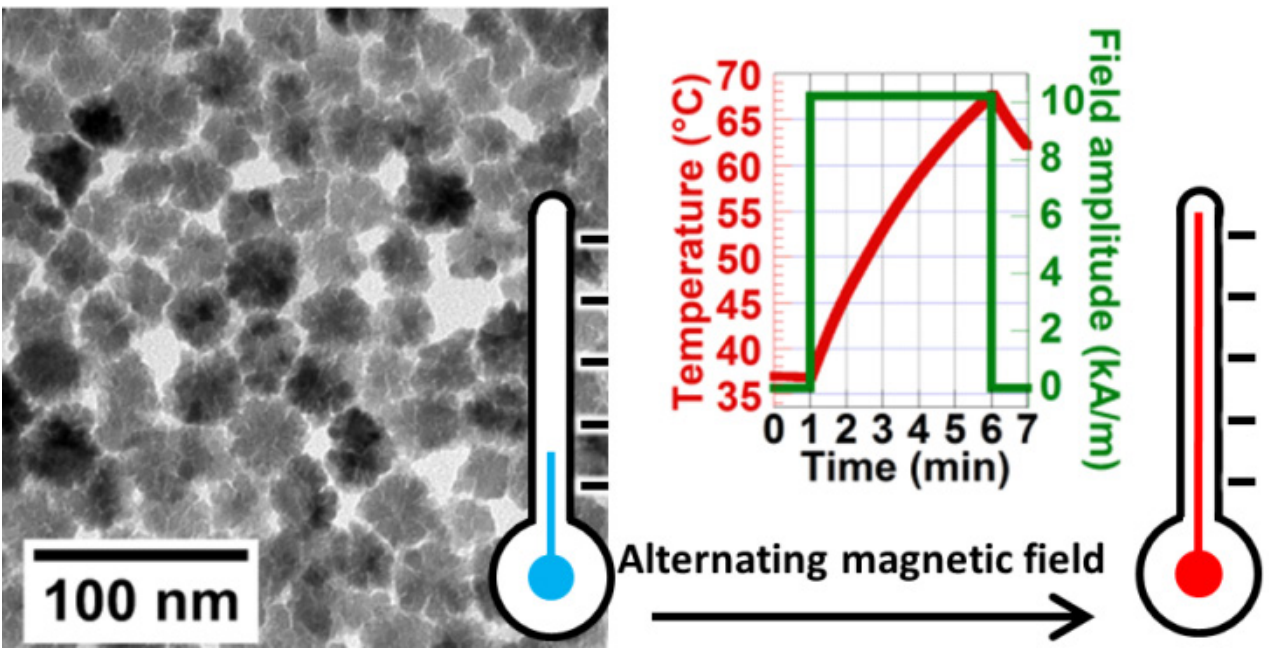

**For Table of Contents Only.** The polyol route offers an excellent compromise between size and shape control and industrial up-scalability for the synthesis of iron oxide nanoparticles. However, this study highlights the critical role played by stoichiometric amounts of water during the forced hydrolysis mechanism, and describes a full library of samples of varying structures (mono-core vs. multi-core) and properties optimized for magnetic fluid hyperthermia and/or MRI contrast.

**SUPPORTING INFORMATION FOR:**

# Tuning sizes, morphologies, and magnetic properties of mono- *vs.* multi-core iron oxide nanoparticles through the controlled addition of water in the polyol synthesis

Gauvin Hemery[a], Anthony C. Keyes Jr.[a], Eneko Garaio[b]*, Irati Rodrigo[b,c], Jose Angel Garcia[c,d], Fernando Plazaola[b], Elisabeth Garanger[a], Olivier Sandre[a]*

[a] LCPO, CNRS UMR 5629/ Univ. Bordeaux/ Bordeaux-INP, ENSCBP 16 avenue Pey Berland, 33607 Pessac, France

[b] Elektrizitatea eta Elektronika Saila, UPV/EHU, 48940 Leioa, Spain

[c] BCMaterials, Parque Tecnológico de Bizkaia, Ed. 50, 48160 Derio, Spain

[d] Fisika Aplikatua II Saila, UPV/EHU, 48940 Leioa, Spain

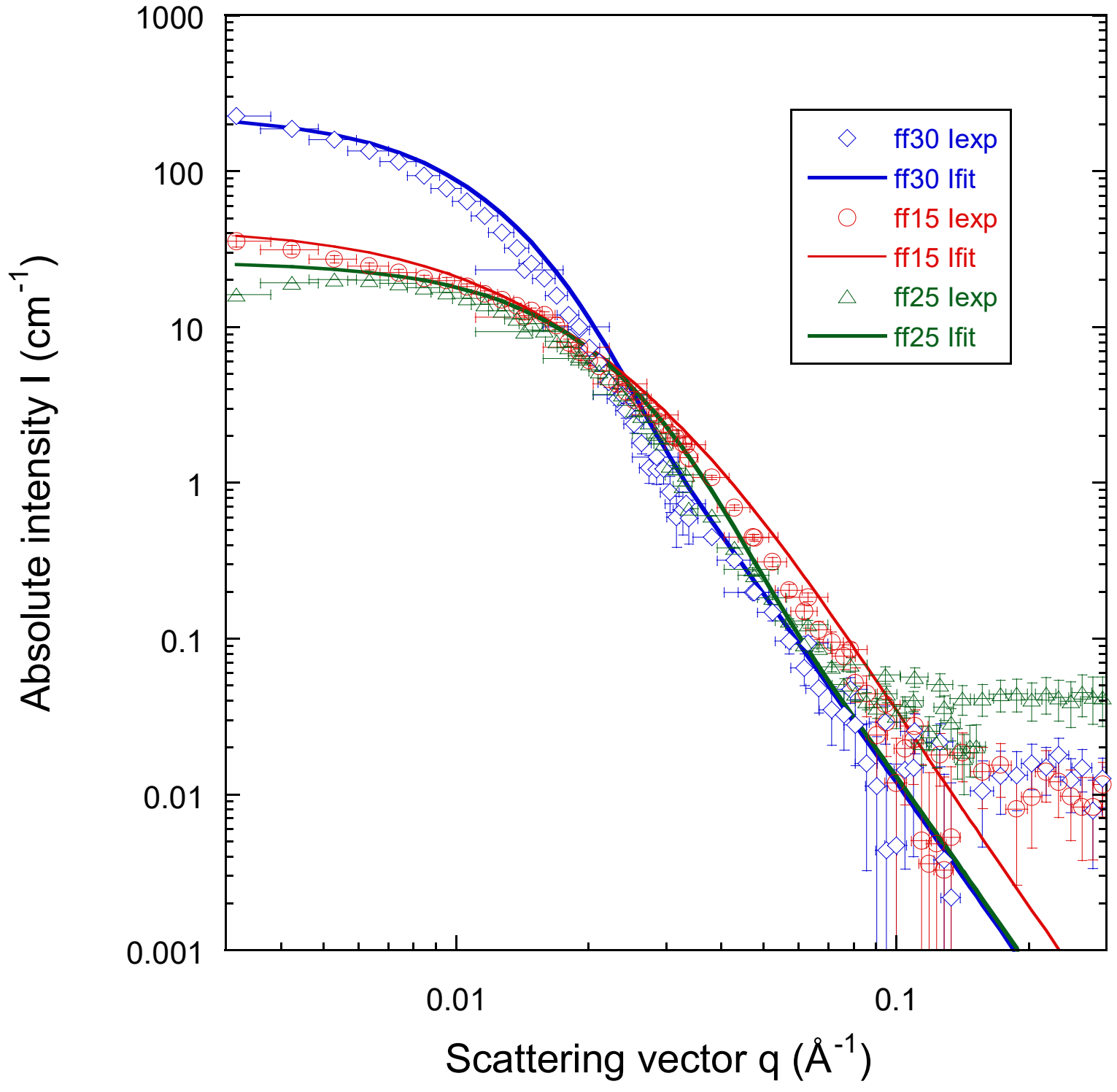

**Figure S1.** Small angle neutron scattering curves of samples 30ff (nanoflowers), 15ff (nanoflowers) and 25ff (smooth spheres). Solid lines represent fits with polydisperse sphere form factor corresponding to following values of the intensity-average radius determined by SANS, to compare with the gyration radius by SANS and the radius obtained by TEM analysis. The specific area was estimated by the Porod law describing the intensity in the high $q$ region: $\left[I(q) - I_{\text{background}}\right] \cdot q^4 = 2\pi\phi(1-\phi)(\Delta SLD)^2 \cdot A_{\text{spe}}$ where $\Delta SLD = 7.53 \cdot 10^{10}\ \text{cm}^{-2}$ stands for the neutron scattering contrast, *i.e.* the difference of scattering length density $(SLD)$ calculated between the IONPs made of $\gamma$-$Fe_2O_3$ $(6.97 \cdot 10^{10}\ \text{cm}^{-2})$ that represent a volume fraction $\phi$, and the aqueous medium $(-5.6 \cdot 10^9\ \text{cm}^{-2}$ for light water). The experimental specific area $A_{\text{spe}}$ $(\text{cm}^2/\text{cm}^3)$ or $A_{\text{spe}}/\rho$ $(\text{cm}^2/\text{g})$ can be compared to the theoretical value calculated for smooth spheres of radius $R_{\text{SANS}}$ and volumetric mass density $\rho$=5 g·cm$^{-3}$, $A_{\text{spe}}^{\text{the}} = 3/R_{\text{SANS}}/\rho$.

30ff: $R_G$=18.9 nm, $R_{\text{SANS}}$=18.1 ± 3.9 nm, $R_{\text{TEM}}$=23.5 ± 4.2 nm, $A_{\text{spe}}$=31.7 m$^2$·g$^{-1}$. $A_{\text{spe}}^{\text{the}}$=33.1 m$^2$·g$^{-1}$.

15ff: $R_G$=19.0 nm, $R_{\text{SANS}}$=10.6 ± 1.8 nm, $R_{\text{TEM}}$=17.5 ± 2.4 nm, $A_{\text{spe}}$=89.9 m$^2$·g$^{-1}$ $A_{\text{spe}}^{\text{the}}$=56.6 m$^2$·g$^{-1}$.

25ff: $R_G$=10.5 nm, $R_{\text{SANS}}$=10.4 ± 2.3 nm, $R_{\text{TEM}}$=7.1 ± 1.6 nm, $A_{\text{spe}}$=52.6 m$^2$·g$^{-1}$ $A_{\text{spe}}^{\text{the}}$=57.7 m$^2$·g$^{-1}$.

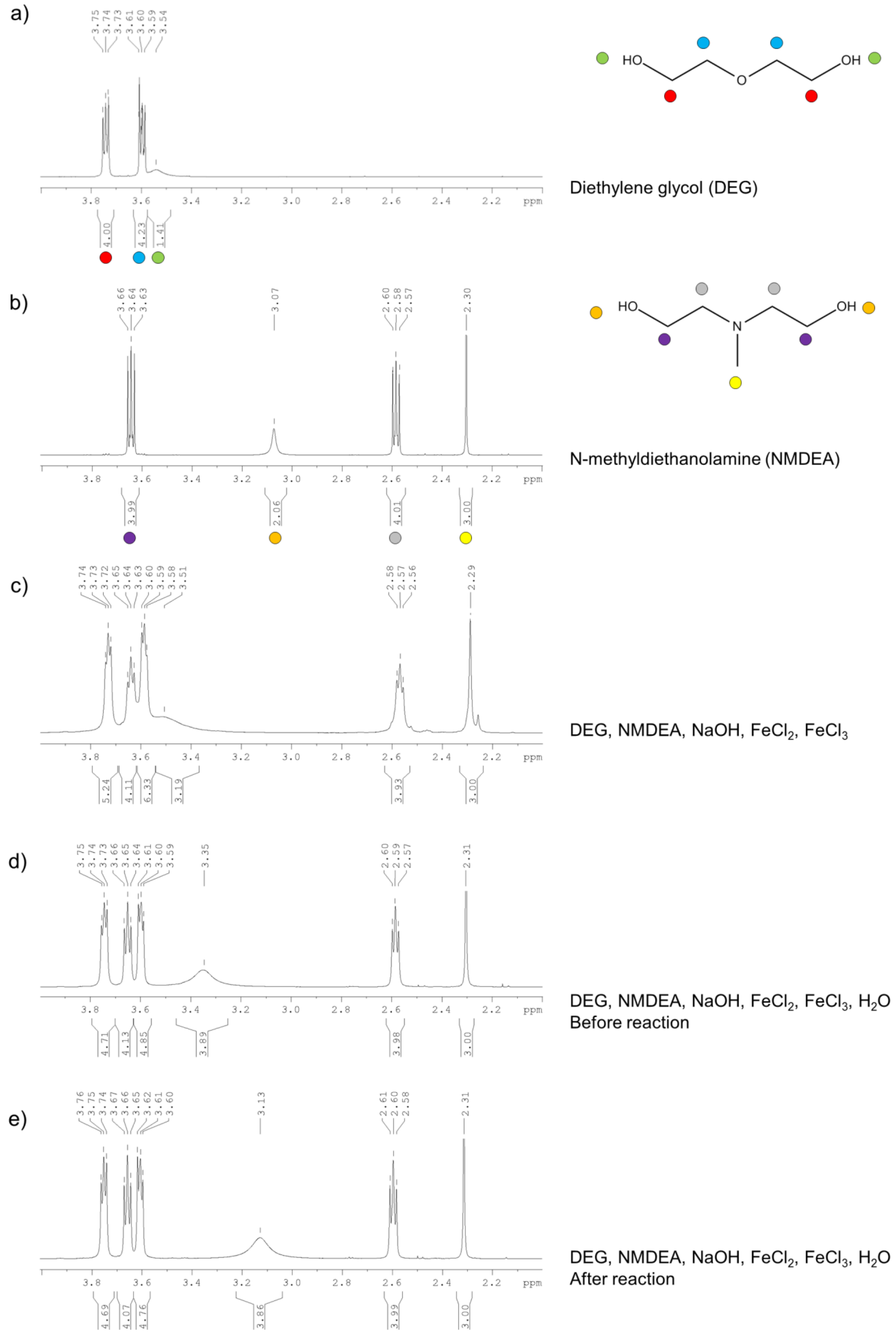

**Figure S2.** NMR analyses of DEG (a), NMDEA (b) and of the medium of synthesis before addition of water (c), after addition of water (d) and after reaction (e). Addition of water shifts the peak corresponding to hydroxyls, which is further shifted after reaction, possibly because of a variation of pH. The solvents do not undergo degradation during the reaction.

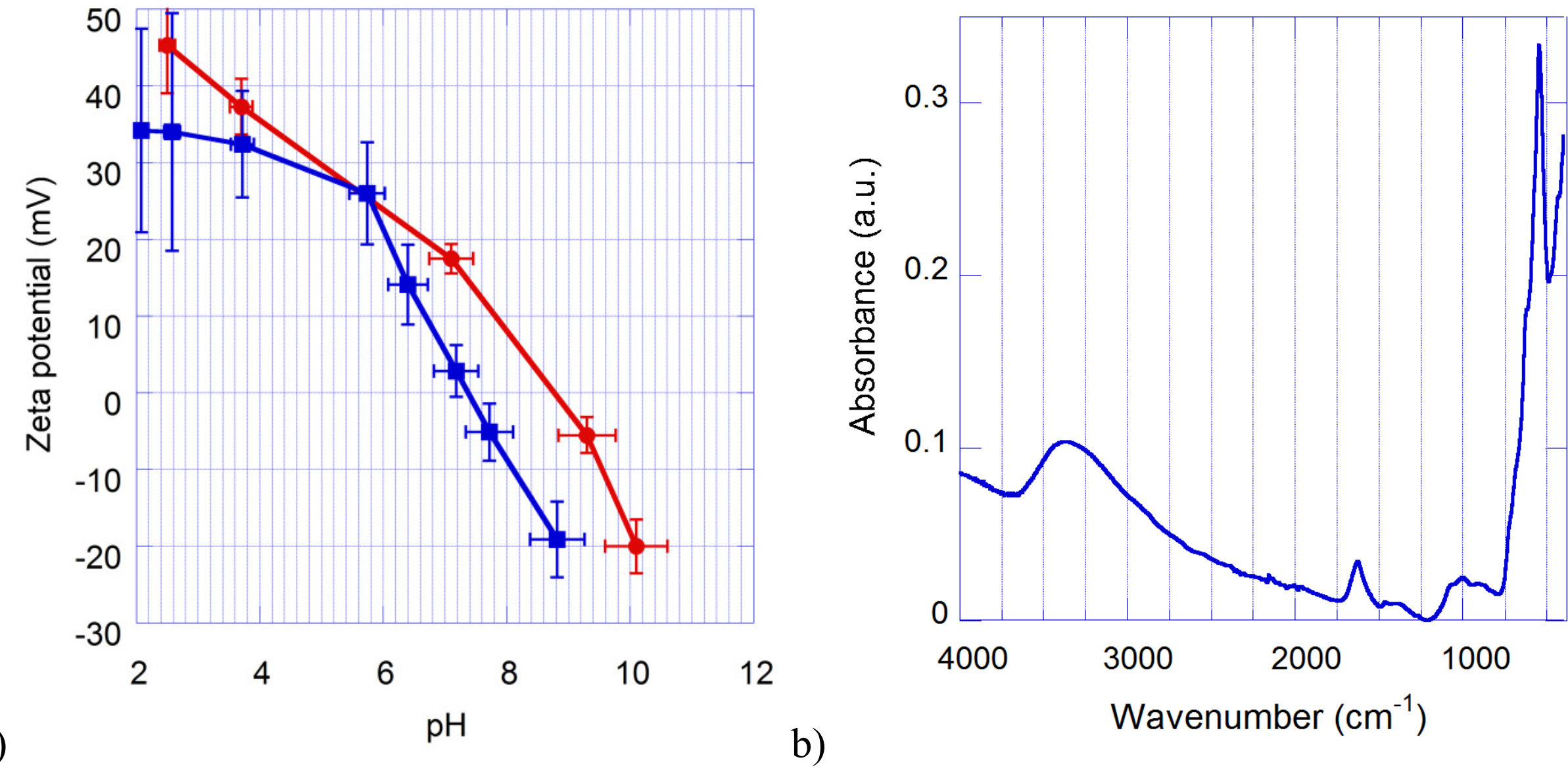

**Figure S3.** a) Zeta (ζ) potential measurement of IONPs whose surface is partially cleaned after the polyol synthesis (red circles), with an isoelectric point (IEP) estimated to be at around pH=9, evidencing the presence of amine moieties of NMEDEA of the solvent molecules chelated on the IONPs. When fully cleaned (blue squares), IONPs show an IEP estimated around pH=7, which is consistent with values previously reported for a bare surface of iron oxide NPs (*i.e.* coated only with neutral hydroxyl groups); b) ATR-IR spectrum of IONPs after washings (batch 21ff) presenting a broad peak in the 3000-3500 $cm^{-1}$ range attributed to stretching vibrations of water. The bending vibration of water produces a peak at 1650 $cm^{-1}$, and the stretching vibration of the bond between oxygen and iron corresponds to the strong peak starting at 540 $cm^{-1}$. This evidences that the material is composed of iron oxide with water adsorbed, but no remaining traces of organic residues, that would yield $CH_2$ bands near 2800-2900 $cm^{-1}$.

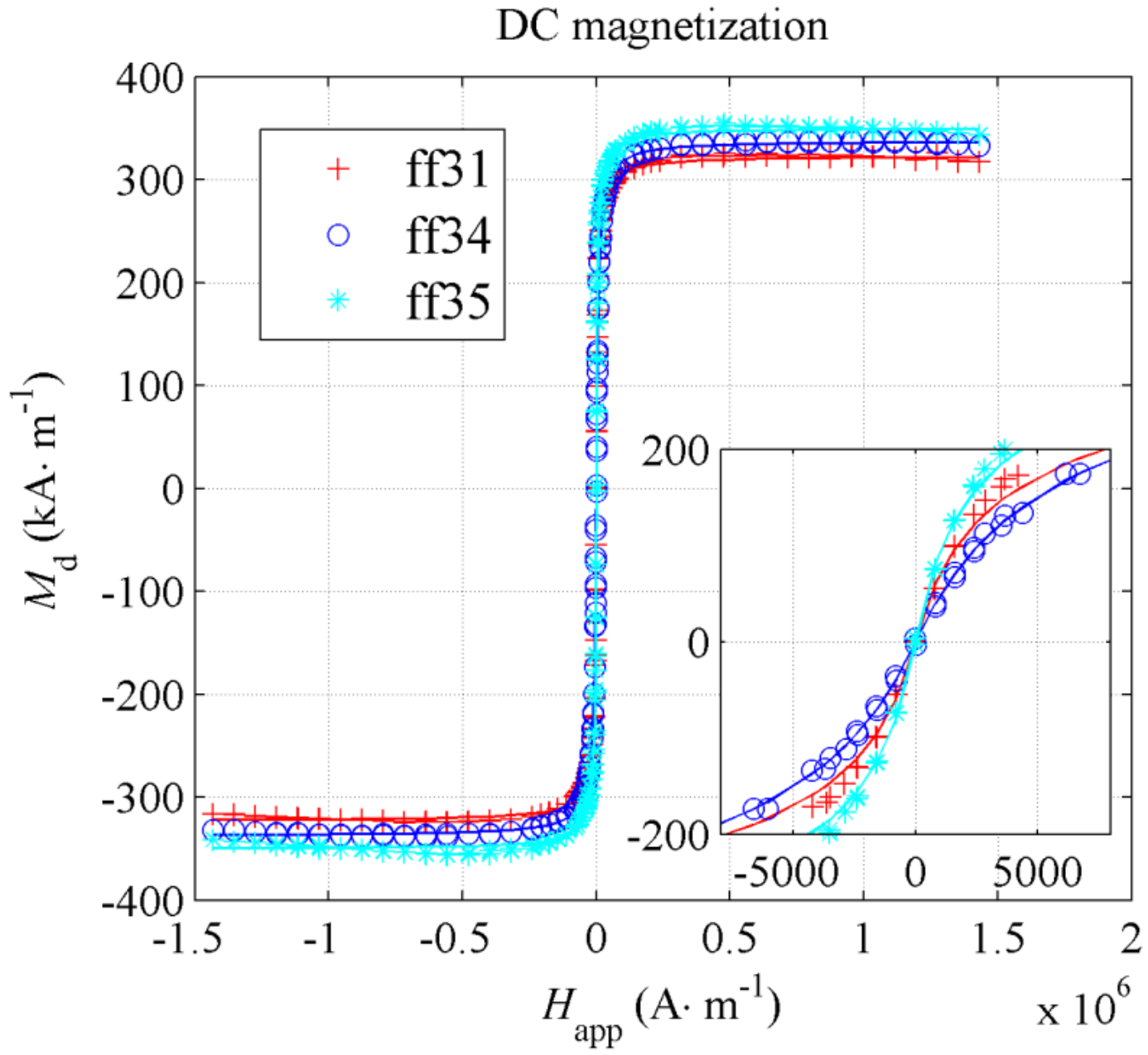

**Figure S4.** DC magnetization of samples 31ff, 34ff and 35ff measured by VSM. As the inset shows, they follow a superparamagnetic behavior. The signal was normalized by volume fraction calculated from the iron oxide concentration (assuming a mass density of 5 $g{\cdot}cm^3$) in order to derive the specific magnetization per domain, denominated $M_d$. In the linear region described by $M_d=\chi{\cdot}H_{app}$, (*i.e.* in a narrow range -1.6< $H_{app}$<1.6 $kA{\cdot}m^{-1}$, the susceptibility value varies between $\chi$=45 (ff34), $\chi$=68 (ff31) and up to $\chi$=100 (ff35).

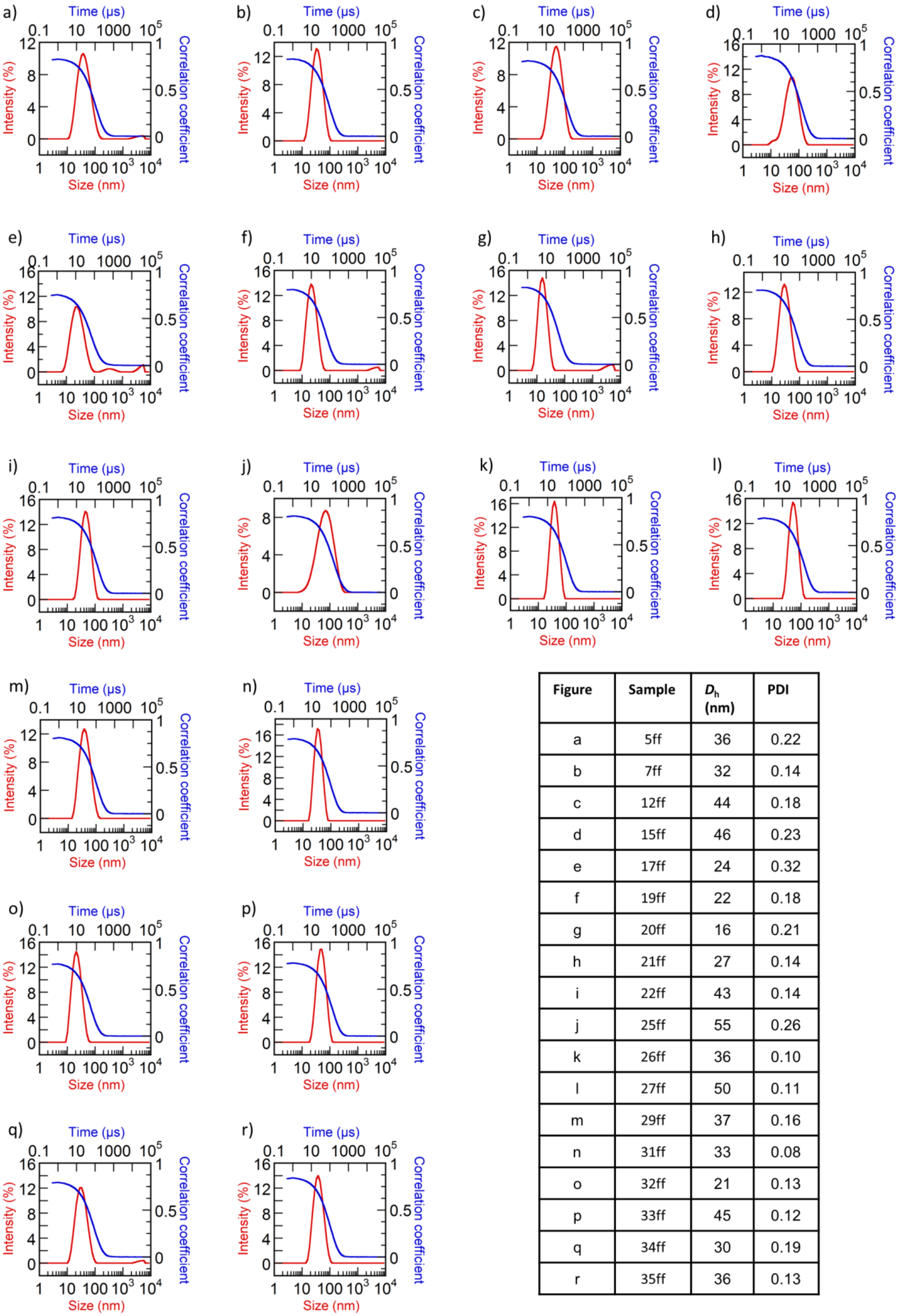

| Figure | Sample | $D_h$ (nm) | PDI |
|---|---|---|---|
| a | 5ff | 36 | 0.22 |
| b | 7ff | 32 | 0.14 |
| c | 12ff | 44 | 0.18 |
| d | 15ff | 46 | 0.23 |
| e | 17ff | 24 | 0.32 |
| f | 19ff | 22 | 0.18 |
| g | 20ff | 16 | 0.21 |
| h | 21ff | 27 | 0.14 |
| i | 22ff | 43 | 0.14 |
| j | 25ff | 55 | 0.26 |
| k | 26ff | 36 | 0.10 |
| l | 27ff | 50 | 0.11 |
| m | 29ff | 37 | 0.16 |
| n | 31ff | 33 | 0.08 |
| o | 32ff | 21 | 0.13 |
| p | 33ff | 45 | 0.12 |
| q | 34ff | 30 | 0.19 |
| r | 35ff | 36 | 0.13 |

**Figure S5.** DLS correlograms and intensity-averaged distributions of diameters. The 2$^{nd}$ order cumulant fit leads to the *Z*-average hydrodynamic diameter ($D_h$)and polydipsersity index (PDI).

**Table S1.** Summary of characteristics of the sample library synthesised by the polyol route: mean diameter $d_0$ and standard deviation $\sigma$ obtained by particle counting on TEM images of $N$>100 IONPs on (figures in parentheses correspond to the inner core size of nanoflowers when it can be measured; an asterisk indicates a lower statistics); the initial temperature slope during 5 s under an AMF $H_{app}$=10.2 $kA \cdot m^{-1}$ at $f$=755 kHz and at 3 $g \cdot L^{-1}$ iron oxide; the specific absorption rate (normalized by the mass of iron oxide); the $Z$-average hydrodynamic diameter ($D_h$) and polydispersity index (PDI), the $r_1$ and $r_2$ relaxivities and their ratio (measured at 37°C and under 1.41 Tesla). *Lines in italics correspond to nanoflowers* (otherwise nanospheres). **Bold lines indicate samples analysed in details in main text**. Missing values correspond to measurements that were not performed because samples were estimated not interesting enough, or because of absence of colloidal stability due to a too large TEM size (*e.g.* 9ff).

| Sample | $d_0 \pm \sigma$ (nm) | $T$ slope ($°C \cdot s^{-1}$) | SAR ($W \cdot g^{-1}$) | $D_h$ (nm) | PDI | $r_1$ ($s^{-1} \cdot mM_{Fe}^{-1}$) | $r_2$ ($s^{-1} \cdot mM_{Fe}^{-1}$) | $r_2/r_1$ |
|---|---|---|---|---|---|---|---|---|
| 4ff | 8.3±1.9 | - | - | - | - | 12.1 | 113.3 | 9.3 |
| *5ff* | *21.4±3* | *0.080* | *111* | *36.3* | *0.22* | *5.7* | *80.3* | *14.0* |
| 7ff | 7.1±1.6* | 0.022 | 31 | 32.0 | 0.14 | 18.3 | 170.6 | 9.3 |
| *9ff* | *60.0±7.4* | - | - | - | - | - | - | - |
| 12ff | 10.0±2 | 0.052 | 72 | 44.2 | 0.18 | 22.0 | 149.6 | 6.8 |
| *13ff* | *11.7±4.7** | *0.14* | *189* | *57.2* | *0.35* | *15.6* | *291.3* | *18.7* |
| ***15ff*** | ***36.9±4.8***<br>***(7.4±1.4)*** | **-** | ***296*** | ***46.3*** | ***0.227*** | **-** | **-** | **-** |
| 16ff | 4.7±1.2 | - | - | 29.3 | 0.39 | 6.4 | 23.9 | 3.8 |
| 17ff | 4.3±1.1 | 0.004 | 6 | 25.2 | 0.36 | - | - | - |
| 18ff | 3.4±0.8 | 0.006 | 8 | 44.2 | 0.96 | 5.5 | 12.1 | 2.2 |
| 19ff | 7.7±1.7 | 0.070 | 98 | 21.5 | 0.18 | - | - | - |
| 20ff | 6.3±1.6 | 0.016 | 22 | 16.2 | 0.21 | - | - | - |
| 21ff | 11.2±1.9 | 0.060 | 84 | 26.9 | 0.14 | 17.9 | 145.3 | 8.1 |
| 22ff | 12.1±3.1 | 0.14 | 189 | 42.8 | 0.14 | 10.6 | 118.5 | 11.2 |
| 25ff | 14.2±3.2 | 0.090 | 125 | 55.2 | 0.26 | 19.7 | 267.7 | 13.6 |
| *26ff* | - | *0.088* | *123* | *35.7* | *0.10* | *16.0* | *219.1* | *13.7* |
| *27ff* | - | *0.11* | *150* | *49.6* | *0.11* | *12.5* | *194.2* | *15.5* |
| *29ff* | *25.9±4.5*<br>*(12.1±1.9)* | *0.13* | *178* | *36.5* | *0.16* | - | - | - |
| ***31ff*** | ***27.5±4.2*** | ***0.19*** | ***268*** | ***33.3*** | ***0.08*** | ***13.1*** | ***341.9*** | ***26.1*** |
| 32ff | 14.5±3.4 | 0.096 | 134 | 20.6 | 0.13 | 25.9 | 203.4 | 7.8 |
| 33ff | 21.6±5.5 | 0.10 | 145 | 44.7 | 0.12 | 5.9 | 140.6 | 23.8 |
| **34ff** | **32.3±5.0** | **0.15** | **206** | **29.9** | **0.19** | **12.2** | **163.1** | **13.4** |
| ***35ff*** | ***29.1±4.4*** | ***0.19*** | ***265*** | ***35.7*** | ***0.13*** | ***6.2*** | ***170.0*** | ***27.3*** |
| *36ff* | *32.3±5.0* | *0.17* | *237* | *71.0* | *0.37* | *0.86* | *22.3* | *25.9* |

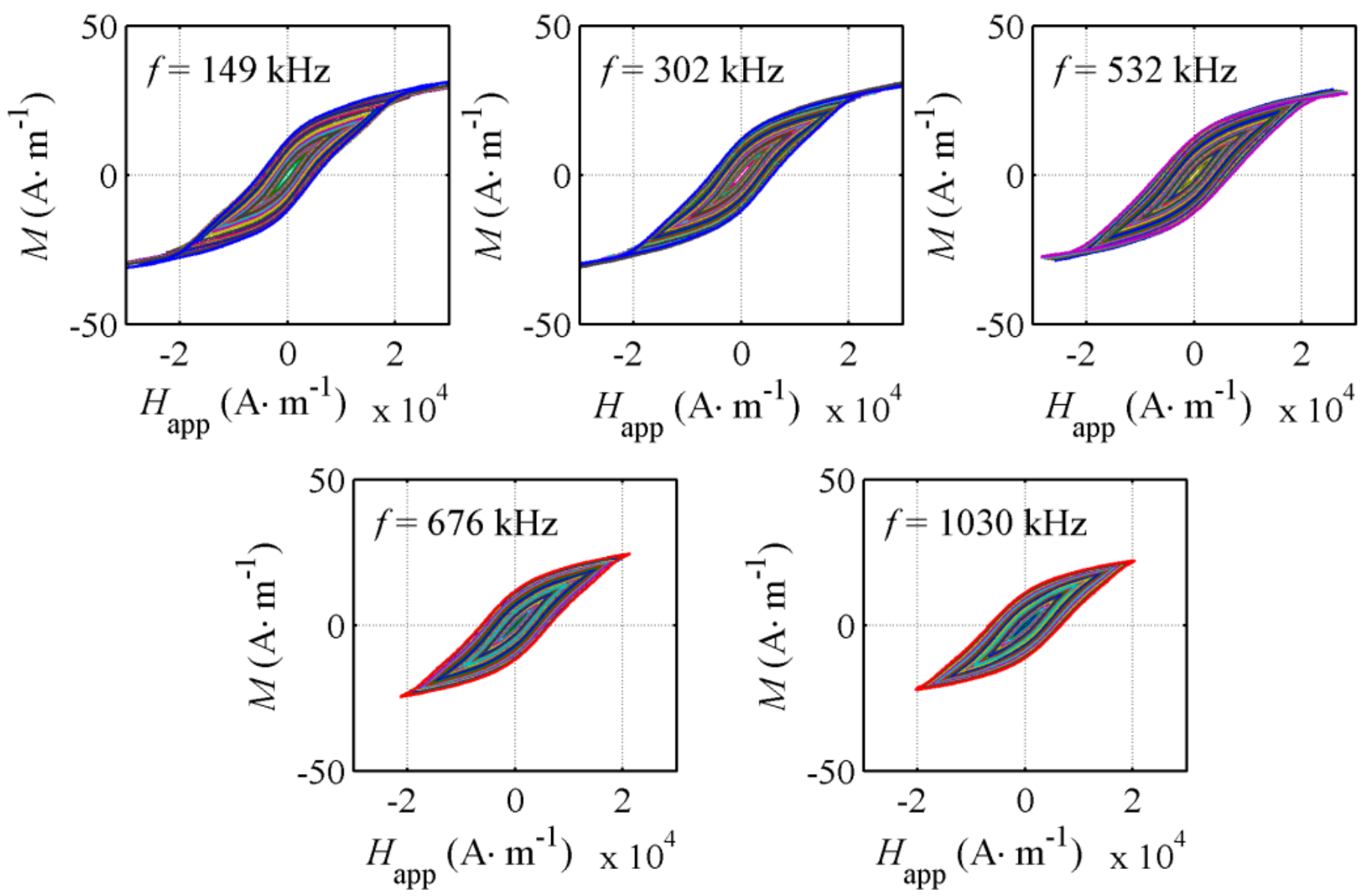

**Figure S6.** AC hysteresis cycles of oxidized 15ff sample (nanoflowers of grain size 7.4±1.4 nm and outer diameter 36.9±4.8 nm) at different AMF amplitudes $H_{app}$ and frequencies (*f*). The superimposed black line is the DC magnetization measured by VSM.

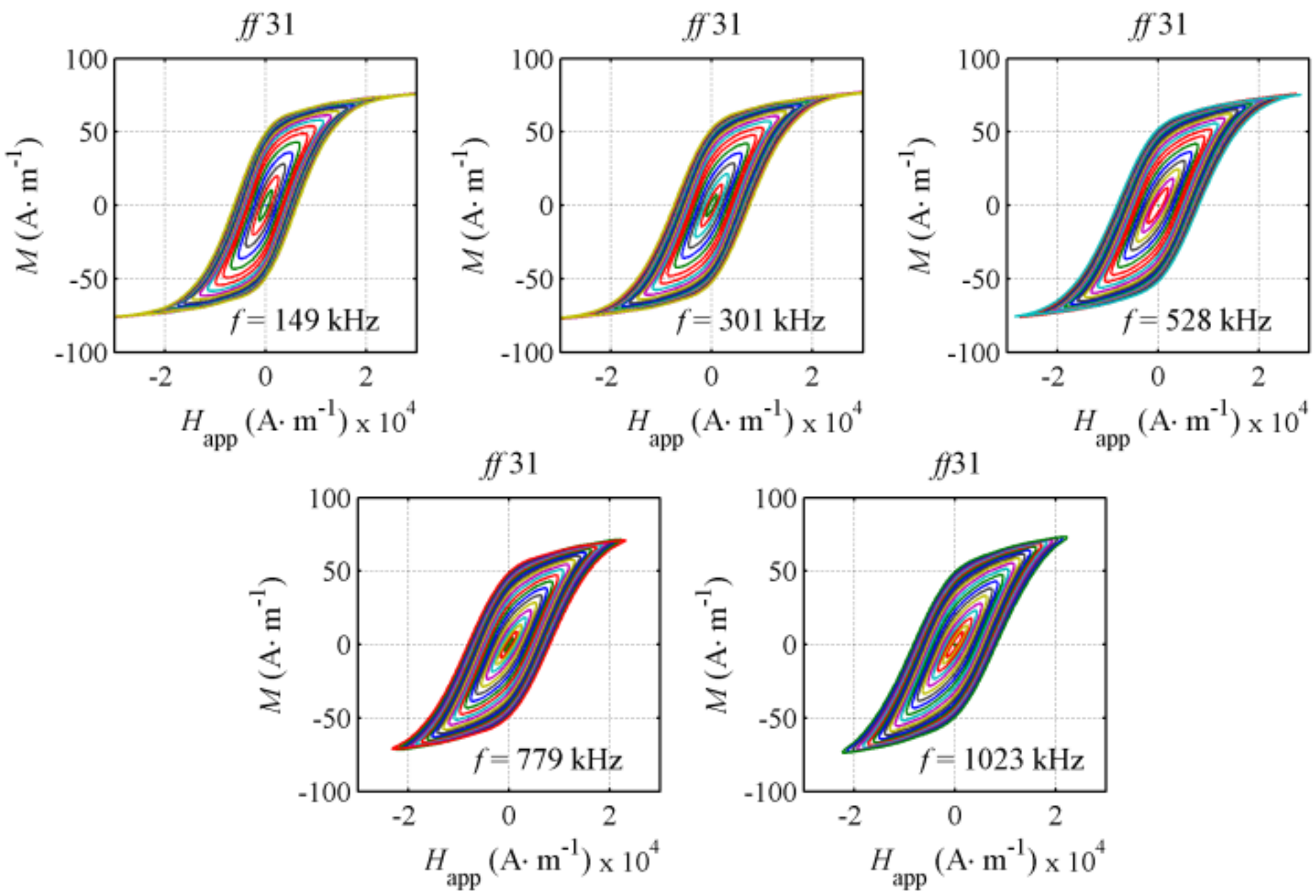

**Figure S7.** AC hysteresis cycles of 31ff sample (nanoflowers of 27.5 ± 4.2 nm outer diameter) at different field amplitudes $H_{app}$ and frequencies (*f*).

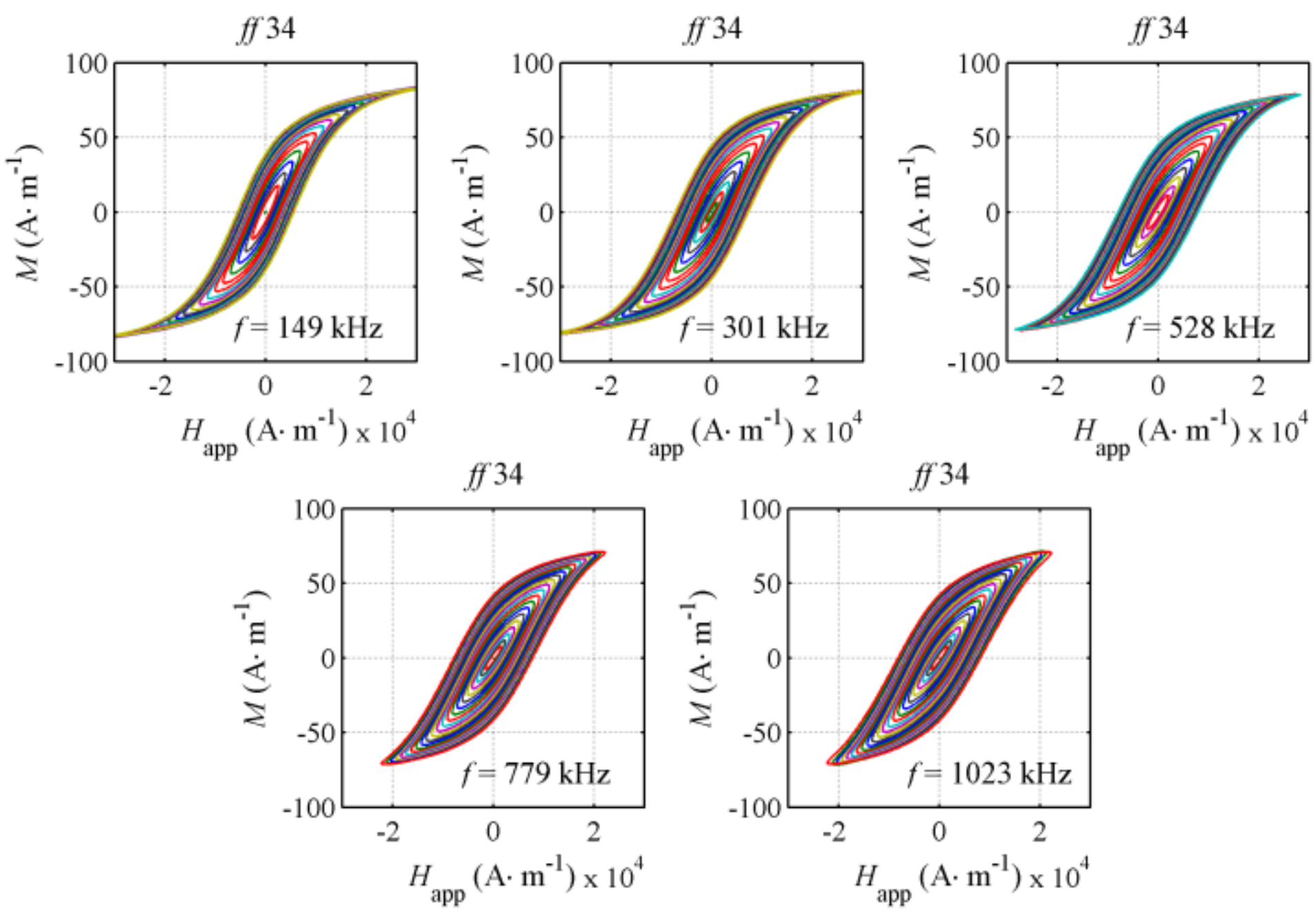

**Figure S8.** AC hysteresis cycles of 34ff sample (smooth nanospheres of 18.5 ± 3.2 nm diameter) at different field amplitudes $H_{app}$ and frequencies (*f*).

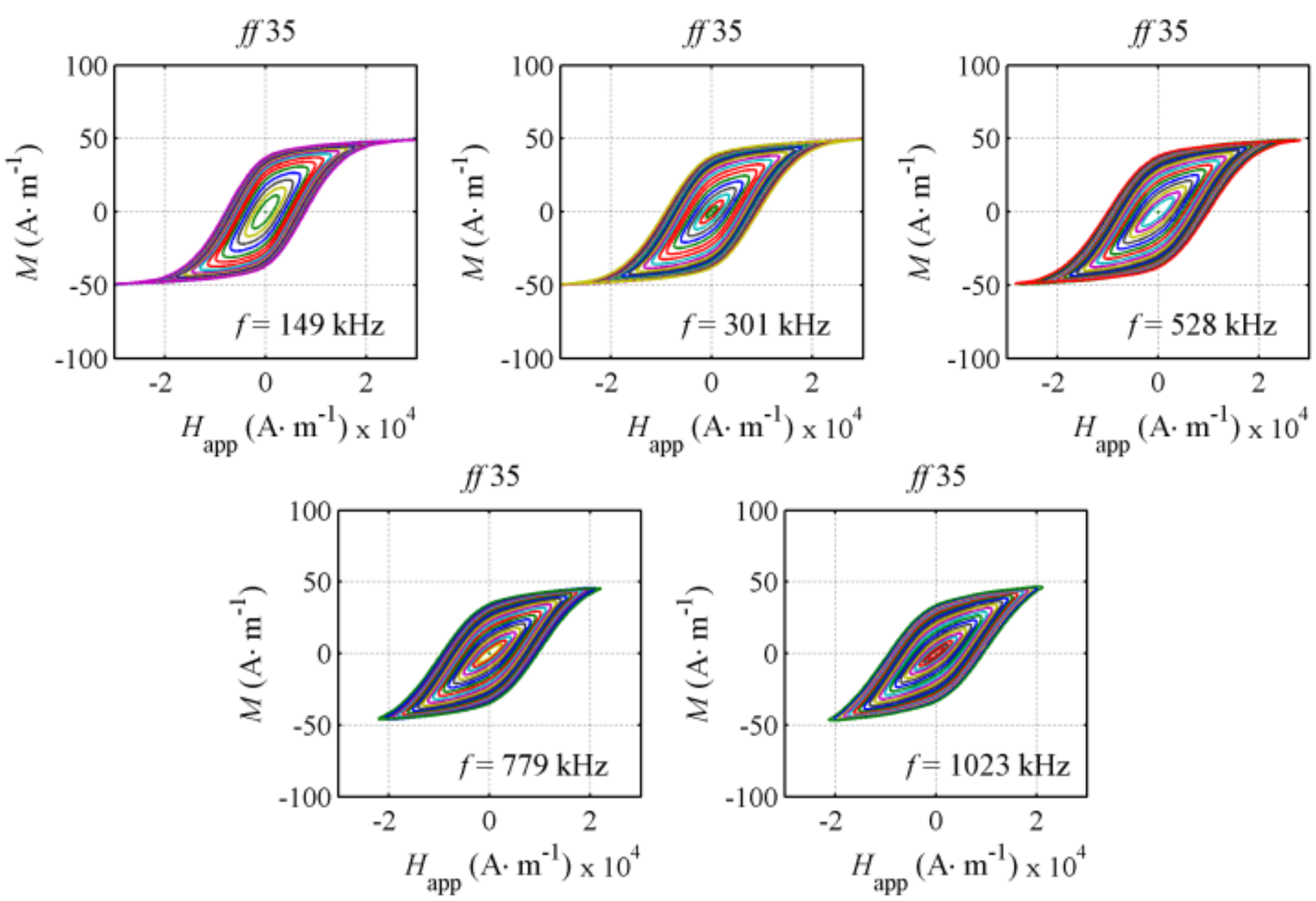

**Figure S9.** AC hysteresis cycles of 35ff sample (nanoflowers of 29.1 ± 4.4 nm outer diameter) at different field amplitudes $H_{app}$ and frequencies (*f*).

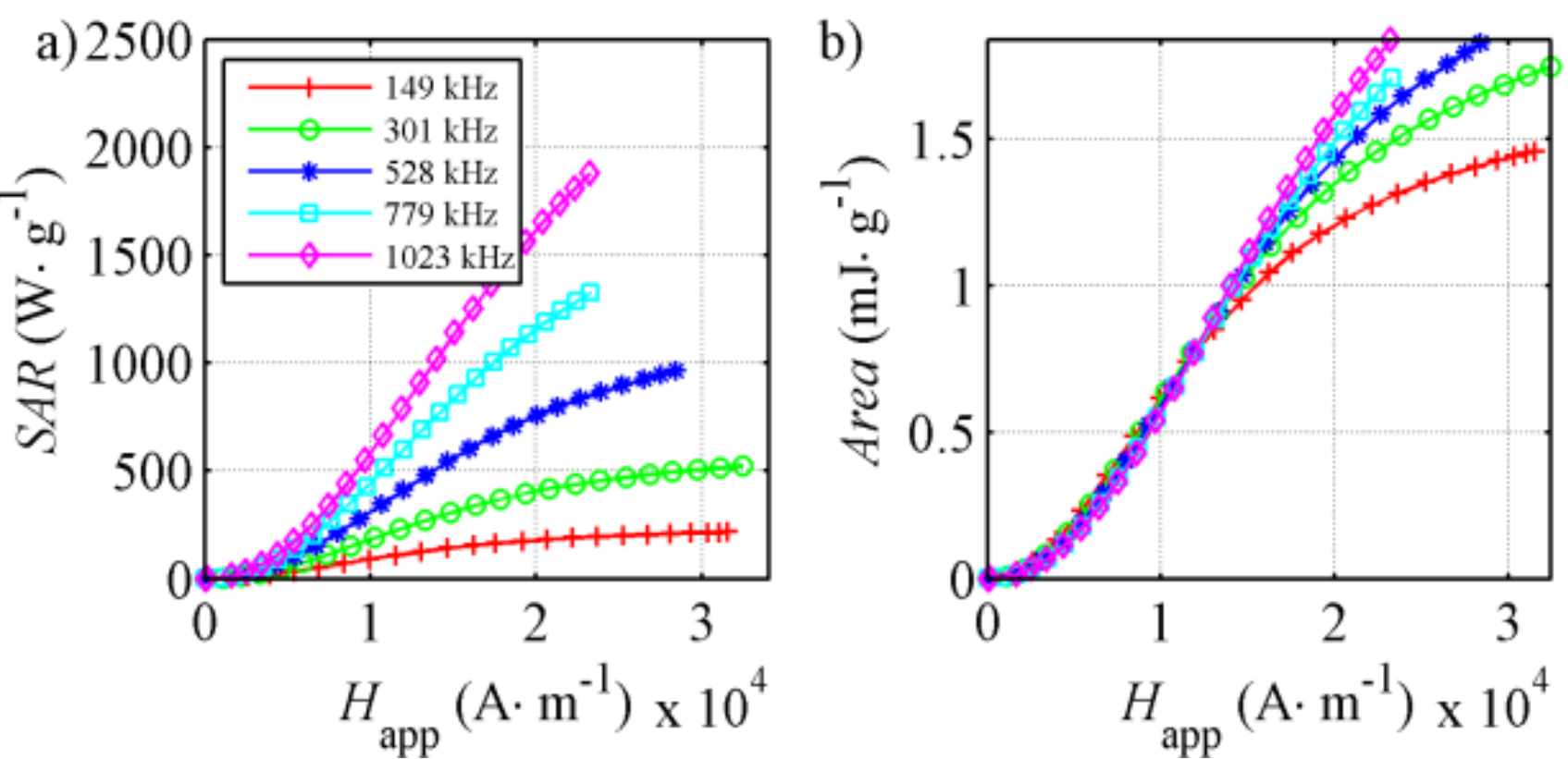

**Figure S10.** *SAR* (a) and hysteresis area (b) of sample 31ff (nanoflowers of 27.5 ± 4.2 nm outer diameter) *versus* applied magnetic field amplitude ($H_{app}$).

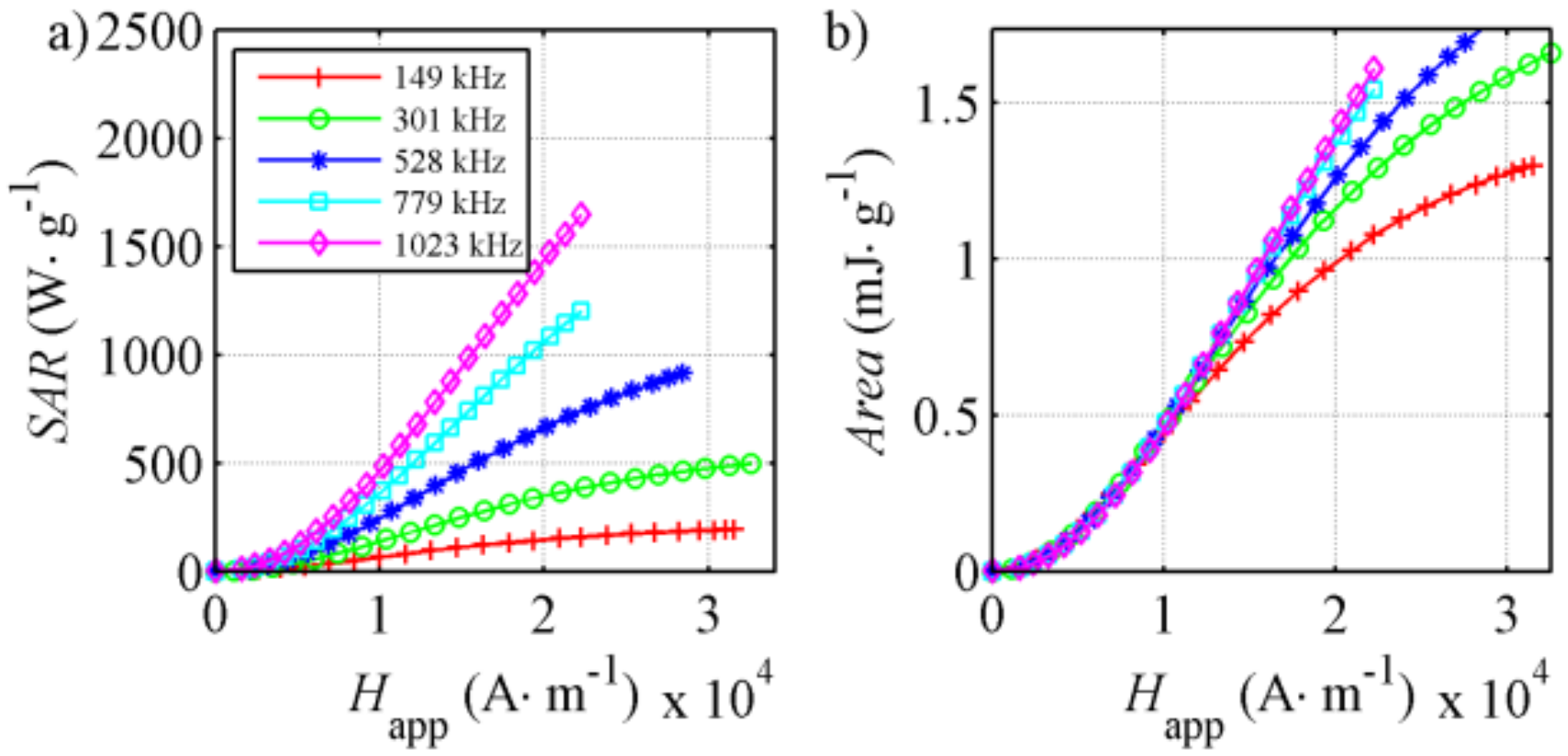

**Figure S11.** *SAR* (a) and hysteresis area (b) of sample 34ff (smooth nanospheres of 18.5 ± 3.2 nm diameter) *versus* applied magnetic field amplitude ($H_{app}$).

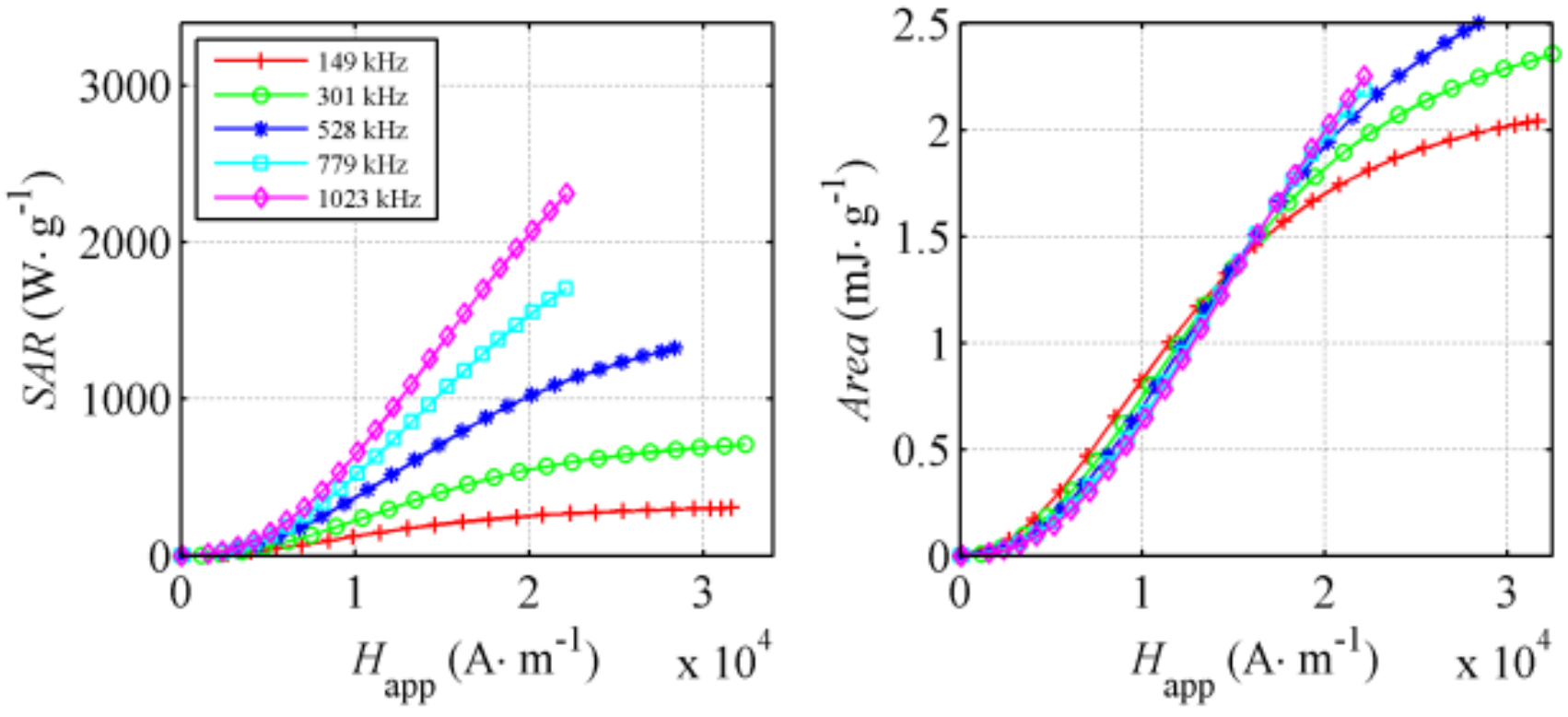

**Figure S12.** *SAR* (a) and hysteresis area (b) of sample 35ff (nanoflowers of 29.1 ± 4.4 nm outer diameter) *versus* applied magnetic field amplitude ($H_{app}$).

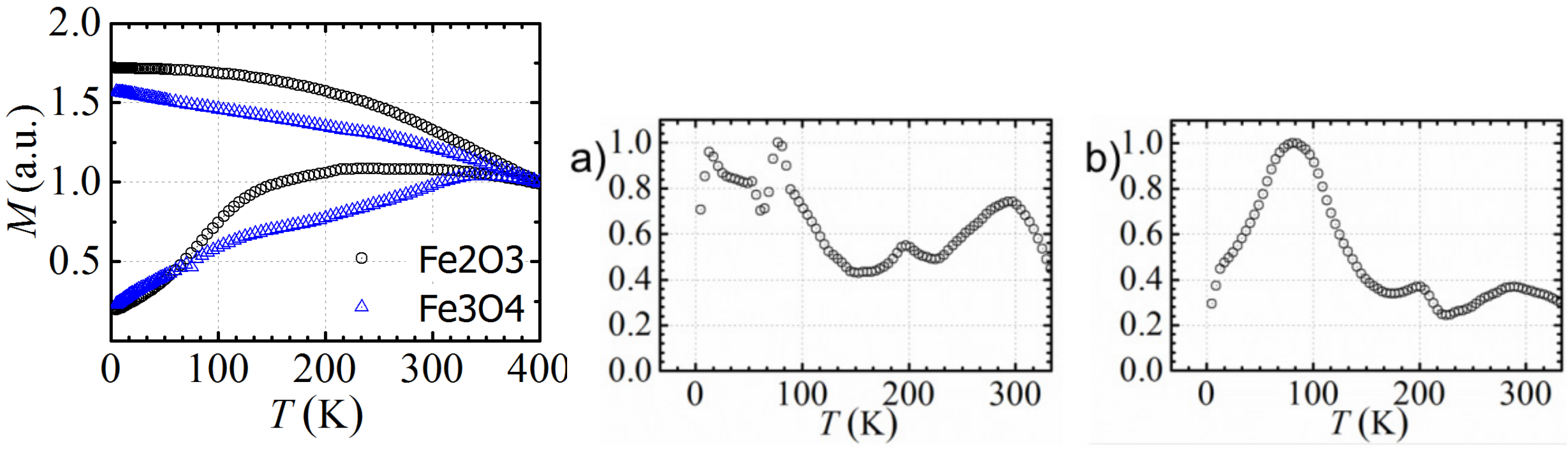

**Figure S13.** ZFC-FC measurement by MPMS magnetometry and derivative d($M_{FC}$-$M_{ZFC}$)/d$T$ of the FC-ZFC curve difference for un-oxidized $Fe_3O_4$ (a) and oxidized γ-$Fe_2O_3$ (b) 15ff nanoflowers. The peak near 90K is ascribed to the Verwey transition.

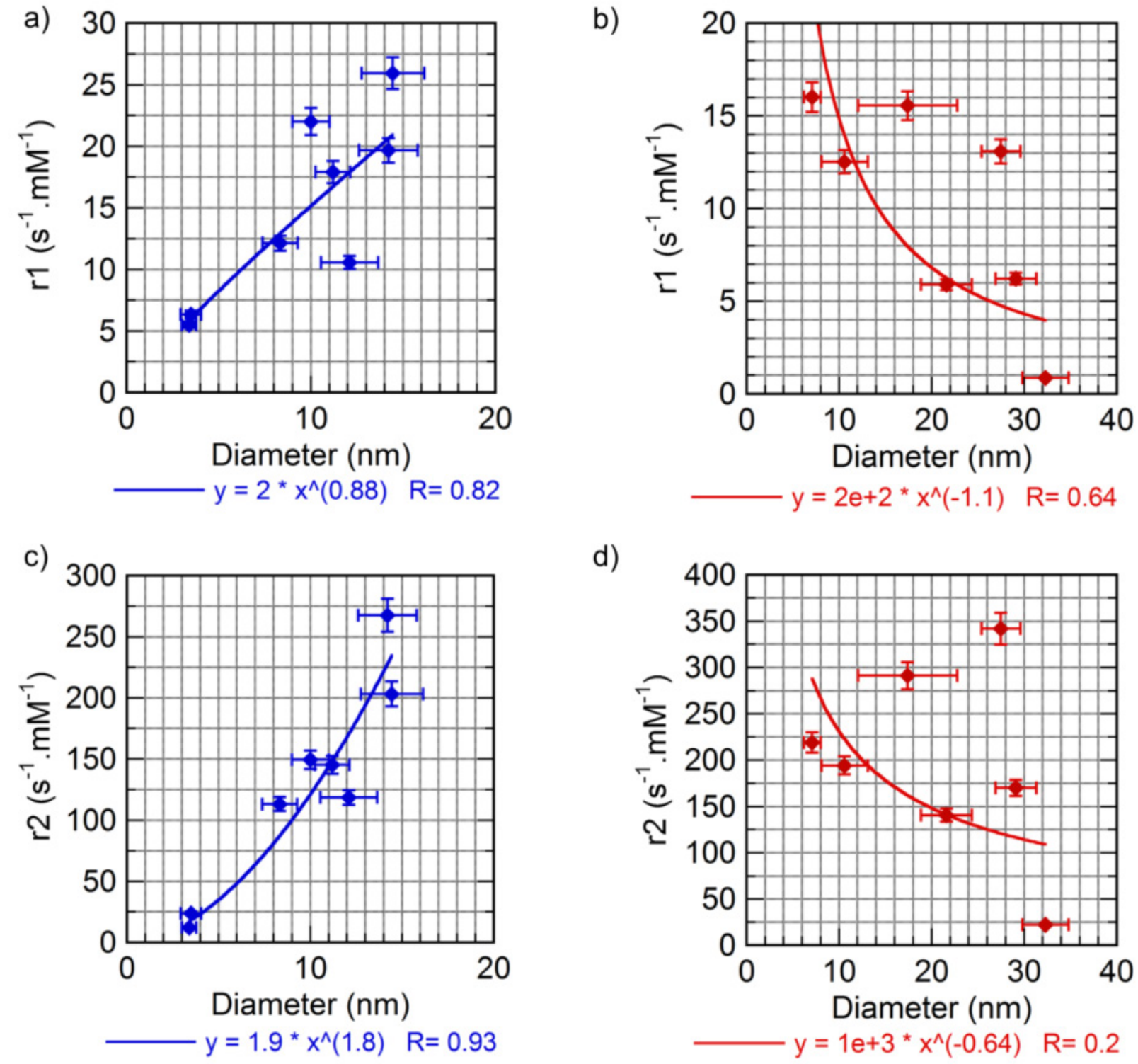

**Figure S14.** Longitudinal relaxivity of smooth spheres (a) and nanoflowers (b). Transverse relaxivity of smooth spheres (c) and nanoflowers (d). All measurements were performed at 37°C on a 1.41 Tesla / 60 MHz Bruker mq60 relaxometer.